\documentclass[aps,pra,twocolumn,showpacs,letterpaper,tighten,float,superscriptaddress,nofootinbib,reprint]{revtex4-1}

\usepackage[utf8]{inputenc}
\usepackage[fleqn]{amsmath}
\usepackage[mathscr]{euscript}
\usepackage{amsfonts}
\usepackage{amssymb}
\usepackage{array}
\usepackage[T1]{fontenc}
\usepackage[margin=1in]{geometry}
\usepackage{lmodern}
\usepackage{tabularx}
\usepackage{fancyhdr}
\usepackage{graphicx}
\graphicspath{{images/}}
\usepackage[utf8]{inputenc}
\usepackage[english]{babel}
\usepackage{amsthm}
\usepackage{braket}
\usepackage{tikz}
\usepackage{subfig, float}
\usepackage{algorithm}
\usepackage{algpseudocode}

\newcommand{\outerp}[2]{\ket{#1}\!\bra{#2}}

\newcommand{\tr}{\text{Tr}}

\newcommand{\mc}{\mathcal}

\newcommand{\mb}{\mathbb}

\newcommand{\intg}{\text{int}}
\newcommand{\orb}{\text{orbit}}

\newcommand{\hyph}{\text{-}}

\begin{document}
\title{A Quantum Solution for Efficient Use of Symmetries in the Simulation of Many-Body Systems}
\author{Albert T. Schmitz}
\email{albert.schmitz@colorado.edu}
\affiliation{
Department of Physics and Center for Theory of Quantum Matter,
University of Colorado, Boulder, Colorado 80309, USA}
\affiliation{Intel Labs, Intel Corporation, Hillsboro, Oregon 97124, USA}
\author{Sonika Johri} 
\affiliation{Intel Labs, Intel Corporation, Hillsboro, Oregon 97124, USA}

\begin{abstract}
A many-body Hamiltonian can be block-diagonalized by expressing it in terms of symmetry-adapted basis states. Finding the group orbit representatives of these basis states and their corresponding symmetries is currently a memory/computational bottleneck on classical computers during exact diagonalization. We apply Grover's search in the form of a minimization procedure to solve this problem.
%, improving upon the existing algorithm to reduce the number of qubits and oracle calls. 
Our quantum solution provides an exponential reduction in memory, and a quadratic speedup in time over classical methods.  We discuss explicitly the full circuit implementation of Grover minimization as applied to this problem, finding that the oracle only scales as polylog in the size of the group, which acts as the search space. Further, we design an error mitigation scheme that, with no additional qubits, reduces the impact of bit-flip errors on the computation, with the magnitude of mitigation directly correlated with the error rate, improving the utility of the algorithm in the Noisy Intermediate Scale Quantum era.
\end{abstract}

\maketitle

\section{Introduction}
As several quantum computing platforms become available for general use, finding practical applications for quantum computers is a key driver for the development and adoption of quantum computing technology. Additionally, since the field is expected to remain in the Noisy Intermediate Scale Quantum (NISQ) era\cite{Preskill2018} for the next few decades, designing error mitigation strategies for these algorithms is essential. In this paper, we identify a new application for quantum computers, as well as show how the algorithm should be implemented in the NISQ era.

Much of the excitement around quantum computing started with the introduction of two algorithms: Shor's factorization algorithm\cite{Shor1997} and Grover's search algorithm\cite{Grover1996}. Though the former represents the paradigmatic example of quantum speed up, the latter has been criticized as often only nominally showing speed-up. The criticism stems from the fact that although the oracle-query scaling is polynomially reduced, any quantum oracle which contains all the information of the database must scale with the size of the database\cite{Mateus2005}. This suggests we must look to problems where the oracle in Grover's search can be applied efficiently, treating it as a means to invert a Boolean function. 

D\"{u}rr and H\o yer\cite{Durr1996} suggested a use for Grover's algorithm as a method to find the minimal element of a database. The general idea is to hold the best-known minimum value and search for a member less than that. If a better value is found, the best-known value is updated and the process is repeated for a set number of oracle calls. Assuming the oracle can be efficiently implemented, such a process might not be ideal in all cases as it still scales exponentially compared to approximation schemes such as adiabatic evolution and related minimization processes such as quantum approximate optimization algorithm (QAOA)\cite{Farhi2014}. However, as the names suggest, these are only approximate methods. Furthermore, adiabatic evolution is sensitive to phase transitions due to a closing gap, and QAOA may require significant classical computational overhead. These limitations ultimately stem from the fact that such methods are sensitive to not just order, but also `distance.' Grover minimization (Gmin) on the other hand is only dependent on the order. It treats the minimum the same whether it's separated from the next largest value by 1 or 100. This suggests that in special cases where an exact minimum is required or where we wish to ignore distance (or there is no notion of distance), Gmin is a good alternative.  

We present one such problem which occurs in the simulation of strongly-correlated materials or quantum chemistry problems where one might perform an exact diagonalization calculation. A many-body Hamiltonian often contains several symmetries which might represent spin symmetries, translation symmetry or various other discrete point-symmetries such as an $n$-fold rotation or reflection. Collectively, these symmetries can be formalized as a discrete group. One can leverage these symmetries by using group representation theory to block-diagonalize the full Hamiltonian\cite{Tinkham2003} in a {\it symmetry-adapted basis}, making the remaining diagonalization computationally cheaper. 

However, to calculate the block-diagonal matrix elements, each of the original basis states must be associated to an orbit representative which, for convenience, is chosen %as the one labeled with the smallest integer value.
by labeling all basis states with a single integer value, and the orbit representative is defined as the element with the smallest integer label. One must also know the group operator connecting a basis state to its representative\cite{Wietek2018}. For large systems, the Hamiltonian cannot be stored explicitly but is calculated on-the-fly during diagonalization which means the matrix elements need to be computed over and over during the computation. For this, one has to either store the representative corresponding to each element in the original basis explicitly, which becomes costly in terms of memory, or calculate them on-the-fly. Thus, finding the orbit representative has become a serious bottleneck for using symmetries in exact diagonalization problems.  For large spin systems, special-purpose hardware such as FPGAs have been considered to ease this bottleneck\cite{FPGA}. In some cases where distributed memory systems are used for the diagonalization calculation, the overhead of the symmetry adapted basis is so large that the authors abandon the symmetry-based approach altogether\cite{Lauchli2011}. A technique for addressing this bottleneck for spin-systems with translational symmetry is proposed in Ref.~\cite{Weibe2013, Wietek2018} using a divide \& conquer method based upon sub-lattice coding. This splits the costs between memory and computational time, but only reduces the time by a constant factor, and the memory by a polynomial amount.

In this paper, we consider the use of Gmin for this problem, which results in a quadratic speed-up over the classical algorithms, and requires virtually no classical memory and relatively little quantum memory. We improve upon the textbook version of Gmin to optimize the number of oracle calls and reduce the number of qubits required to implement the oracle. Furthermore, we show that for many reasonable problem instances, the oracle is poly-log in the size of the group and dimension of the Hamiltonian's Hilbert space assuming the group action generators can be efficiently simulated on a quantum computer, making this a practical use for Grover's algorithm. We consider the full circuit implementation for a benchmark case as well as the effects of error on the performance of the algorithm. Our error-mitigation scheme based on real-time post-selection on measurement results between coherent steps of the algorithm represents a near-term use for pre-fault tolerant quantum computing. Furthermore, using Gmin as a sub-routine in classical exact diagonalization is an example of the power of interfacing quantum and classical machines for hybrid algorithms. Alternately, we envision that this algorithm could also be used as a sub-routine which generates the matrix entries of a larger quantum algorithm using symmetry-adapted basis states to simulate a strongly-correlated quantum system.

The remainder of the paper is structured as follows: In Section II, we introduce the problem of finding the orbit representative, give an overview of the existing classical solutions, and then describe in detail our quantum algorithm, including the full circuit description and an analysis of the running time of the algorithm. Section III shows results from the simulation on the Intel Quantum Simulator\cite{qHIP}. Section IV discusses our error mitigation strategies in the presence of noise and their numerical simulation. %and Section~V discusses the simulations of realistic parameters for near-term hardware. 
We conclude in Section V.

\section{Overview of the Problem and the Quantum Solution}
We first briefly review symmetry-adapted basis states and how their matrix elements are calculated following Refs.~\cite{Tinkham2003,Wietek2018}: For a given many-body problem instance, Let $H$ be the Hamiltonian. We then  characterize its symmetries by operators $g \in G$, such that
\begin{align}
[H,g]=0.
\end{align}
From the group and its associated representation theory, we define the symmetry-adapted basis states as
\begin{align}
\ket{v_\alpha}\propto \sum_{g\in G} \chi(g)_\alpha^* g\ket{v},
\end{align}
where $\alpha$ indexes some one-dimensional %\red{What does one-dimensional mean in this context?} 
representation of $G$, $\chi_\alpha(g)$ is the character of the $\alpha^{th}$ representation evaluated at $g$ and $\ket{v}$ are the original ``position'' basis states, such that the action of $g$ on the basis states is $g \ket{v} = \ket{gv}$. One can see that two symmetry-adapted basis states $\ket{v_\alpha}, \ket{u_\alpha}$ are equal (once normalized) so long as $\ket{u} \in \text{orbit}(\ket{v})$, where $\text{orbit}(\ket{v})$ is the set of all basis elements connected to $\ket{v}$ by a group element. Therefore each block of the Hamiltonian in this basis is characterized by just the representation index, with the states in each block represented by unique orbits, so we can choose a single representative $\ket{\tilde v}$ for each orbit. For simplicity, let's assume the group action is free, which is to say $gv =v$ if and only if $g$ is the identity element. Then all states have the same normalization constant up to phase, $\mc N$. Since $\sum_{g\in G} \chi_\alpha (g) \chi_\alpha^*(g)= |G|$, we find that $\mc N=\frac{1}{\sqrt{|G|}}$. We can now calculate the matrix elements of $H$ for a given block via
\begin{align}
&\braket{\tilde v_\alpha | H |\tilde u_\alpha} \nonumber \\
&= \frac{1}{|G|}\sum_{g_1, g_2 \in G} \chi_\alpha(g_1) \chi_\alpha^*(g_2) \braket{\tilde v | g_1^{-1} H g_2|\tilde u}\nonumber \\
=& \frac{1}{|G|}\sum_{g\in G}\sum_{g_2\in G} \chi_\alpha(g_2 g^{-1}) \chi_\alpha^*(g_2) \braket{g\tilde v |H| \tilde u} \nonumber \\
=& \sum_{g\in G} \chi_\alpha(g) \braket{g^{-1} \tilde v|H| u} \nonumber \\
=& \sum_{v \in \text{orbit}(\tilde v)} \chi_\alpha(g_v) \braket{v|H|\tilde u}.
\end{align}
where we have used the fact that all member of $G$ commute with the Hamiltonian, $\chi_\alpha (g)$ is a one-dimensional representation of the group and define $g_v$ such that $g_v v = \tilde v$. As we can see, one needs $g_v$ to calculate the appropriate character. In practice, one calculates the action of $H$ on the representative state $\ket{\tilde u}$, then sorts all coefficients of the resulting vector according to the orbits to form the appropriately weighted sum for each orbit. If the group action is not free, then one also has to calculate and store the normalization factors which also enter the sum. 

In the rest of this section, we describe the problem of finding the group orbit representative with some comments on the classical methods which are used to solve it. We then propose a quantum method based on Gmin which exponentially reduces the memory cost while yielding a quadratic reduction in computational time.
%\noteSJ{Make sure all Figures are referenced in the text.}

\subsection{Orbit Representative Problem Statement}

With the above motivation, we formally state the orbit representative problem.

Problem statement: suppose we have some finite group $G$ with a group action $ G \times V \to V$ such that $(g,v) \mapsto g v$.  We shall refer to $V$ as the position set and its members positions, though they may not correspond to physical position, but rather index some basis set for a Hamiltonian's Hilbert space. Furthermore, we have some function $\intg(v)$ which totally orders the set $V$. We assume $\intg$ maps to the integer value used to label $v$\footnote{What follows could be mapped to more exotic orderings including partial orders if the phase comparator discussed below can be generalized to the given ordering efficiently.}. Define the orbit of a position  $\orb(v) =\{gv: \text{for all } g\in G\}$, which is represented by $\tilde v \in \orb(v)$ such that for all $u \in \orb(v)$, $\intg(\tilde v) \leq \intg(u)$, i.e. it is the smallest element.  

{\bf Given a member $v\in V$, find the orbit representative $\tilde v$ as well as the group element which gives that representative, i.e find $g_v$ such that $g_v v= \tilde v$.}

 Note that based on the application of this problem from the last section, a near-minimum value for the orbit representative is not sufficient; we need the exact minimum. In the general case, one expects that $\log|G| \ll \log|V| \leq |G| \ll |V|$. Table \ref{tbl} gives a list of the solutions to this problem including Gmin and compares the costs. We denote the classical time complexity cost of computing the group action on an arbitrary member of $V$ by $C(G)$ and in general, the quantum time-complexity cost of implementing an operator $A$ on a quantum computer as $\mc C(A)$.

\begin{table*}[t]
\begin{ruledtabular}
\begin{tabular}{ l  l  l  l }

{\bf Method} & {\bf Cl. Mem} & {\bf Q Mem} & {\bf Time} \\
\hline\\
Look-up & $\mc O(|V|)$ & 0 & $\mc O (\log|V|)$ \\
On-the-fly & $\mc O(1)$  & 0 & $\mc O (|G| C(G))$ \\
Divide \& conquer & $ \mc O (\sqrt{|V|})$ & 0 & $ \mc O (|G| C(G) \log|V|)$ (smaller constant coefficient than on-the-fly)\\
Gmin & $\mc O(1)$ & $\mc O(2 \log|V| + \log |G|)$ &$ \mc O\left(\sqrt{|G|}\left( \mc C(\hat G)+ \text{polylog}(|V|\right)\right)$ 

\end{tabular}
\caption{List of the different methods for solving the group representative problem.  We generally expect that $\log|G| \ll \log|V| \leq |G| \ll |V|$. $C(G)$ is the classical cost to calculate the action of $G$ on an arbitrary $v$, while $\mc C(G)$ is the quantum cost to calculate the action of $G$ on an arbitrary $v$.}\label{tbl}
\end{ruledtabular}
\end{table*}

\subsection{Classical Solutions}

There are three classical means of addressing this problem: 
\begin{enumerate}
 \item Look-up: Store orbit representatives corresponding to every element in $V$ and connecting group elements in a look-up table. This can then be efficiently searched when needed, but it requires $\mc O(|V|)$ amount of memory.

\item On-the-fly: When needed, calculate the full orbit to find the smallest element and the connecting group element. This is efficient in terms of memory, but the computation scales as $\mc O(|G|)$.

\item Divide \& conquer: There exist sub-lattice coding methods \cite{Wietek2018}, which allow one to split the costs between memory and computation (see Table~\ref{tbl} for these costs).
\end{enumerate}

While the divide \& conquer method represents a significant reduction in the resources needed, this bottleneck can still be prohibitively expensive. To the best of our knowledge, no one has considered using quantum methods for solving this problem as we discuss in the next section.

\subsection{ Overview of the Grover Minimization Algorithm}

In this section we look to use the Gmin algorithm to solve the problem. We first review the algorithm as given in Ref. \cite{Durr1996} and then adapt it for this problem which includes modifications to optimize the memory and time costs.

Gmin utilizes the function $f_v: G \to V$ such that $g \mapsto f_v(g)=\intg(g v)$ acting on an unsorted database of $|G|$ items; $g$ acts as an index and we want to find the index which points to the smallest value in $f_v$. To encode the group, we also introduce an index on the group elements $g: \mb N_{<|G|} \to G$ such that $x \mapsto g(x)$\footnote{For notation convenience, we equivocate $f_v$ with $f_v \circ g$ and we mean the latter throughout the remainder of the paper.}. Then the number of bits (qubits) needed to index all members of the group is $m = \mc O(\log |G|)$. The original algorithm proceeds as follows:

\medskip
Let $\alpha$ be some real, positive number which we refer to as the {\it oracle budget parameter} from which we define $\alpha \sqrt{|G|}$ as the {\it oracle budget}. Using two quantum registers each of size $m$ (referred to as the group registers), choose an index $0<y<|G|-1$ randomly, and repeat the following, using no more than $\alpha \sqrt{|G|}$ Grover steps:

\begin{enumerate}
 \item Initialize the two registers in the state $\left(\frac{1}{\sqrt{|G|}} \sum_x \ket{x} \right)\ket{y}$,

\item Mark all  $x$ in the first register such that $f_v(x)< f_v(y)$,

\item Apply a ``Grover search with an unknown number of marked elements'' (Gsun) \cite{Boyer1998} to the first register and

\item measure the first register with outcome $y'$; if $f_v(y')<f_v(y)$, $y \gets y'$.
\end{enumerate}

It is argued in the reference that for $\alpha = \frac{45}{2}$, the second register holds the minimum value with a probability of at least $50\%$. Below, we discuss how to relate the success rate and $\alpha$ using numerical methods. Appendix~\ref{derive} gives a modified analytic derivation such that one finds a better value of $\alpha=\frac{45}{8}$ to achieve a success rate of at least $50 \%$. 

To make this algorithm more explicit, we must address how to implement the second and third steps, which is equivalent to a method for implementing Gsun and its oracle. In general for Grover search, if the number of marked elements is known, one can apply the exact number of Grover steps to reach one of the marked states with high probability. However, this probability is not monotonic with the number of oracle calls. One can ``overshoot'' the target state and reduce the probability of reaching the answer with additional oracle calls. Thus, not knowing the number of marked elements could be problematic if we don't include some additional procedures. We refer those unfamiliar with Grover's search algorithm to Refs.~\cite{Grover1996,Nielsenbook} for details. Ref.~\cite{Boyer1998} provides a solution given by Gsun. Gsun iterates the search and randomly draws the number of Grover steps from a running interval. Those authors prove that the probability of selecting a marked element is asymptotically bounded below by $\frac{1}{4}$, thus insuring we can find a marked element with probability greater that 50\% after a number of oracle calls that still scales as $\sqrt{|G|}$. 

\begin{figure}
\centering
\includegraphics[scale=.7]{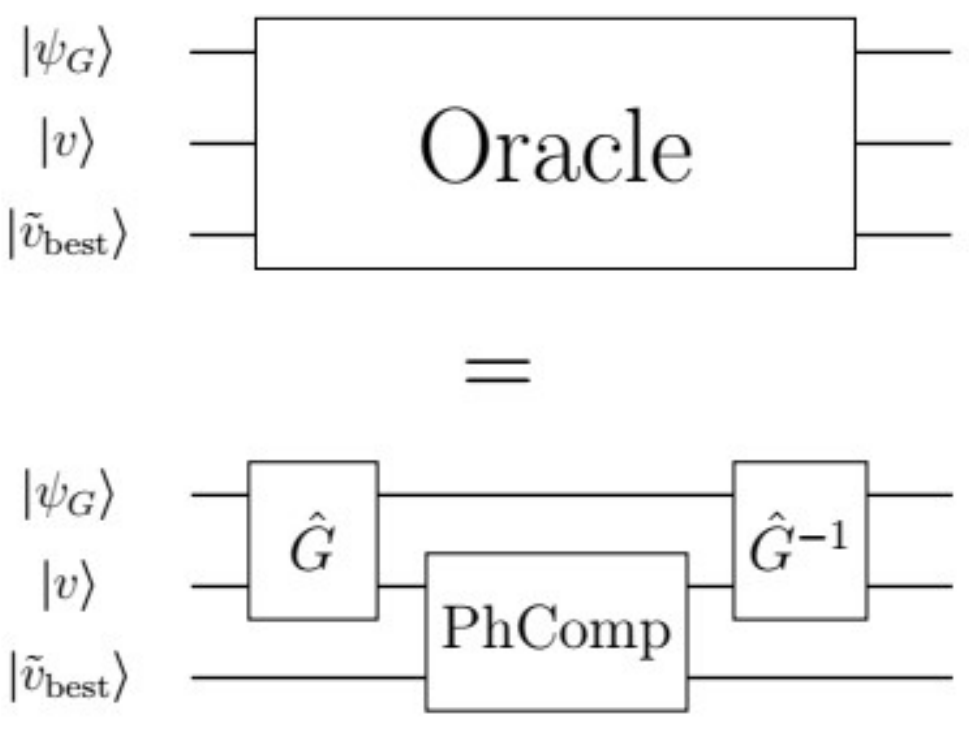}
\caption{Circuit diagram for our proposed oracle.}\label{oracle}
\end{figure}

\begin{figure}
\centering
\includegraphics[scale=.7]{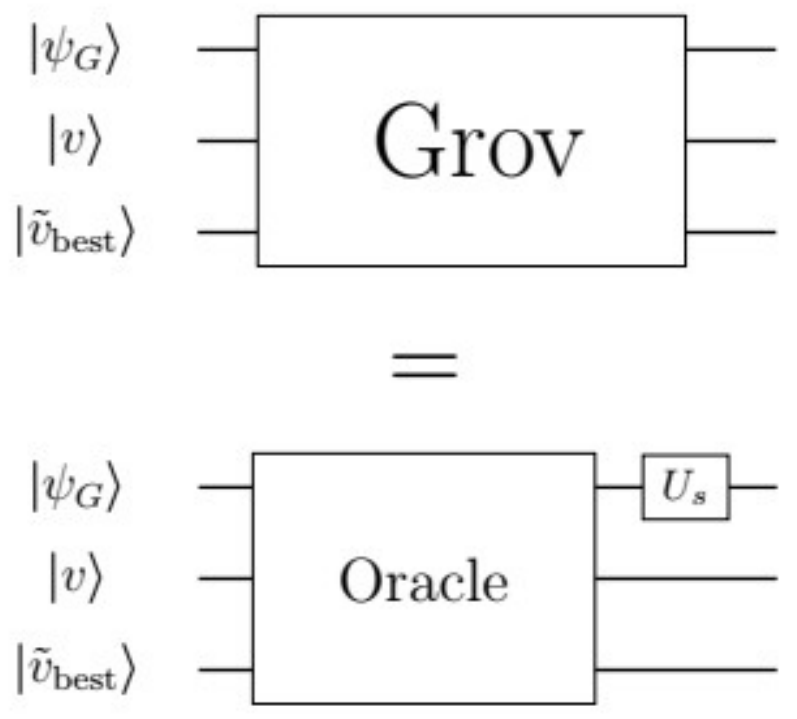}
\caption{Circuit diagram for Grov.}\label{Grov}
\end{figure}

To mark elements as in step two, we must define the oracle. According to Refs.~\cite{Boyer1998, Grover1996}, marking an element means the oracle produces the action on any computational basis state $\ket{x}$,
\begin{align}
\text{Oracle} \ket{x} = \begin{cases}
-\ket{x} & \text{ if $x$ is marked,}\\
\ket{x} & \text{ otherwise}
\end{cases}.
\end{align}
Note the second step requires we calculate $f_v(x)$ and $f_v(y)$ which implies we also require quantum registers to hold these values. There may exist multiple methods for implementing such an oracle, but the simplest  and perhaps cheapest method for our problem is to further hold the value $v$ in a quantum register of size $n =\mc O(\log |V|)$ which we refer to as the first {\it position register}. Furthermore, we replace the second group register with a second  position register of size $n$. So our method is not to store the best-known value for the group index ($y$ in the above algorithm) as was done in previous implementations of Grover minimization, but rather store $\tilde v_{\text{best}}=f_v(y)$ in a quantum register. $y$ can then be stored classically and updated when $\tilde v_{\text{best}}$ is updated. This innovation reduces the number of gates and qubits required for the oracle. The oracle is then implemented as follows: We first implement the group action operator $\hat G$ on the group register and the first position register which has been initialized with $v$ such that
\begin{align}
\hat G \ket{x}\ket{v} = \ket{x}\ket{g(x) v}.
\end{align}
We then apply a quantum circuit that in general acts on two quantum registers of equal size such that it applies a negative sign to the state if the computational basis state of the first register is less than that of the second. We refer to this circuit as {\it phase comparator} (PhComp)  which has the behavior
\begin{align}\label{eq:phcomp}
\text{PhComp} \ket{a} \ket{b}= \begin{cases}
- \ket{a}\ket{b} & \text{ if $a<b$}, \\
\ket{a} \ket{b} & \text{ otherwise}
\end{cases}.
\end{align}
So after applying the group action operator, we apply PhComp to the two position registers, and then uncompute the group action operator. This completes the oracle as show in Fig. \ref{oracle}. To complete one Grover step (Grov), we then apply the usual reflection operator defined as
\begin{align} \label{eq:ref}
U_s= I- 2 \outerp{s}{s} = V \left( I-2\outerp{0}{0}\right) V^\dagger,
\end{align}
where $\ket{s}= \frac{1}{\sqrt{|G|}}\sum_{x} \ket{x}$ and $V$ is any unitary such that $V\ket{0} = \ket{s}$. For completeness, the circuit for Grov is shown in Fig. \ref{Grov}. 

%As another modification, the set number of oracles (oracle budget) is given by $\frac{45}{2} \sqrt{|G|}$ in accordance with the upper bound established in Ref. \cite{Durr1996}. In the appendix, we preform a more careful calculation to find that this bound can be tightened to $\frac{45}{8} \sqrt{|G|} \approx 5.6 \sqrt{|G|}$. This is a considerable savings in the runtime for the algorithm, especially for near-term implementation, and the numerical study below suggest this can be tightened further $\sim 4.5 \sqrt{|G|}$.

 If we unpack Gsun and integrate this into our modified version of Gmin, the psuedo-code flow of the algorithm is shown in Algorithm 1.

%\bigskip
%{\scriptsize
%START
%
%\begin{itemize}
%\item[]Allocate a quantum register $\ket{\psi_G}$ of size $m$  \hs \#allocate quantum memory
%\item[]Allocate a quantum register $\ket{\psi_1}$ of size $n$
%\item[]Allocate a quantum register $\ket{\psi_2}$ of size $n$  
%\item[]
%\item[]$\tilde v_{\text{best}} = v$
%\item[]$x_{\text{best}}= 0$
%\item[]
%\item[] While $cnt < \alpha \sqrt{|G|}$ \hs \# combines Gmin and GSun to stop after a fix number of calls
%\begin{itemize}
%\item[]$p=$rand$(0, t-1)$
%
%\item[]$cnt= cnt + p$
%\item[]
%\item[]Initialize $\ket{\psi_G} \to \ket{0}$ \hs \# initialize quantum registers
%\item[]Initialize $\ket{\psi_1} \to \ket{v}$
%\item[] Initialize $\ket{\psi_2} \to \ket{\tilde v_{\text{best}}}$
%\item[]
%\item[]  $H^{\otimes m} \ket{\psi_G}$  \hs \# apply Grover search
%\item[] $\text{(Grov)}^p \ket{\psi_G}\ket{\psi_1}\ket{\psi_2}$ 
%\item[] Measure$(\ket{\psi_G} \to x)$
%\item[]
%\item[] If $f_v(x)<\tilde v_{\text{best}}$ \hs \# compare and update best-known values
%\begin{itemize}
%\item[]$\tilde v_{\text{best}}= f_v(x)$
%\item[]$ x_{\text{best}} = x$
%\item[]$ t=\max(1, \beta t)$ \hs \# this ends the call to Gsun
%\end{itemize}
%\item[] else
%\begin{itemize}
%\item[]$t= \min (\gamma t, \sqrt{N})$
%\end{itemize}
%\item[]
%\end{itemize}
%\item[] return $x_{\text{best}}, \tilde v_{\text{best}}$
%\end{itemize}
%
%END

\begin{algorithm}[H]
\caption{Grover Minimization} \label{alg1}
\begin{algorithmic}[1]
\State Allocate QRegister $\ket{\psi_G}$ of size $m$
\State Allocate QRegister $\ket{\psi_1}$ of size $n$
\State Allocate QRegister $\ket{\psi_2}$ of size $n$
\State $v_{\text{best}}=\gets v, x_{\text{best}}\gets 0, c\gets 0, t\gets 1$  
 \While{ $c< \alpha \sqrt{|G|}$} 
\State $p\gets $rand$(0,t-1)$\;
 \State $c \gets c +p +1$\;
\State Initialize $\left(\ket{\psi_G} \otimes\ket{\psi_1} \otimes\ket{\psi_2} \gets \ket{0}\otimes\ket{v} \otimes \ket{v_{\text{best}}}\right)$\;
%\State Initialize $\ket{\psi_1} \gets \ket{v}$\;
%\State Initialize $\ket{\psi_2} \gets \ket{\tilde v_{\text{best}}}$ \;
 \State $ V \ket{\psi_G}$\;
\State$\text{(Grov)}^p \ket{\psi_G}\ket{\psi_1}\ket{\psi_2}$ \;
\State Measure$(x \gets \ket{\psi_G})$\;

  \If{ $f_v(x)< v_{\text{best}}$}
 \State $v_{\text{best}}\gets f_v(x)$\;
\State $ x_{\text{best}} \gets x$\;
 \State $ t\gets \max(1, \beta t)$\;
\Else
 \State $t\gets \min (\gamma t, \sqrt{|G|})$\;
\EndIf
 \EndWhile
\State \Return $v_{\text{best}}, x_{\text{best}} $
\end{algorithmic}
\end{algorithm}

%On the line ending the call to Gsun, we have included the condition that $ t=\max(1, \beta t)$. At the start of Gsun, $t=1$, in which case $\beta=0$. But Gsun assumes we know nothing about the number of marked elements which is not strictly true in this case. We know that the number of marked elements has decreases from one call to the next. Thus we choose a $\beta \leq 1$ to reduce the ``ramp-up time'' to reach the critical stage of Gsun ( a concept defined in the reference). For now, this value is empirically determined to be $\beta^{-1}=\sqrt{8}$.  
Note that we have chosen the initializing best guess $v_{\text{best}} =v$,  as we assume $v$ is effectively random. Also, we count the check step in line 12 as an effective oracle call so that the classical and quantum solutions can be more accurately compared. $\gamma  \in \left(1, \frac{4}{3}\right)$ and $ \beta \in \left[0,1\right]$ are additional parameters which we use to minimize $\alpha$. $\gamma$ is discussed in Ref.~\cite{Boyer1998} and controls the rate of the exponential ``ramp-up'' for the parameter $t$ which in turn determines the ceiling of the random sampling for number of oracle calls used in the Grover search step of the algorithm. In principle, a large $\gamma$ reduces the time to reach $t \sim \sqrt{|G|}$ which is optimal if $v_{\text{best}}$ is near the minimum (the number of marked elements is small; the search takes longer). However if $v_{\text{best}}$ is far from the minimum, $\gamma$ being too large and $t\sim \sqrt{|G|}$ increases the chances that we apply too many oracle calls and dramatically overshoot a state of high overlap with a marked element. Thus, we need to balance the rate at which $t$ increases by optimizing $\gamma$. $\beta$ is a parameter which we introduce here. As the algorithm was originally written, after a better value of $v_{\text{best}}$ is found in line 13 of Algorithm \ref{alg1}, Gsun effectively ends and on the next cycle is re-called. Gsun then assumes it knows nothing about how close we are to the minimum by resetting the value of $t$ back to $1$ (as would be the case for $\beta=0$). However, we do know something, namely that we are closer to the minimum than the iteration before (the number of marked elements has decreased). Thus we don't need the ramp-up time for $t$ which is only included to address when we are far from the minimum. By including the $\beta$ parameter, we are looking to exploit this limited knowledge about the number of marked elements. We discuss the exact values chosen for these parameters in Sec.~\ref{sec:obandps}.

\subsection{Circuit Implementation of Grov and its Cost}\label{sec:circ} 

We now discuss a full circuit implementation of all subroutines of Grov. As $\hat G$ is specified by the problem instance, we only give an explicit implementation for the group $G_{\text{add}}^{N}$ which represents addition modulo $N=2^n$ or translation on a cycle of $N$ positions. Otherwise, we discuss a general strategy for more complicated realistic groups. 

\begin{figure}
\centering
\includegraphics[scale=.7]{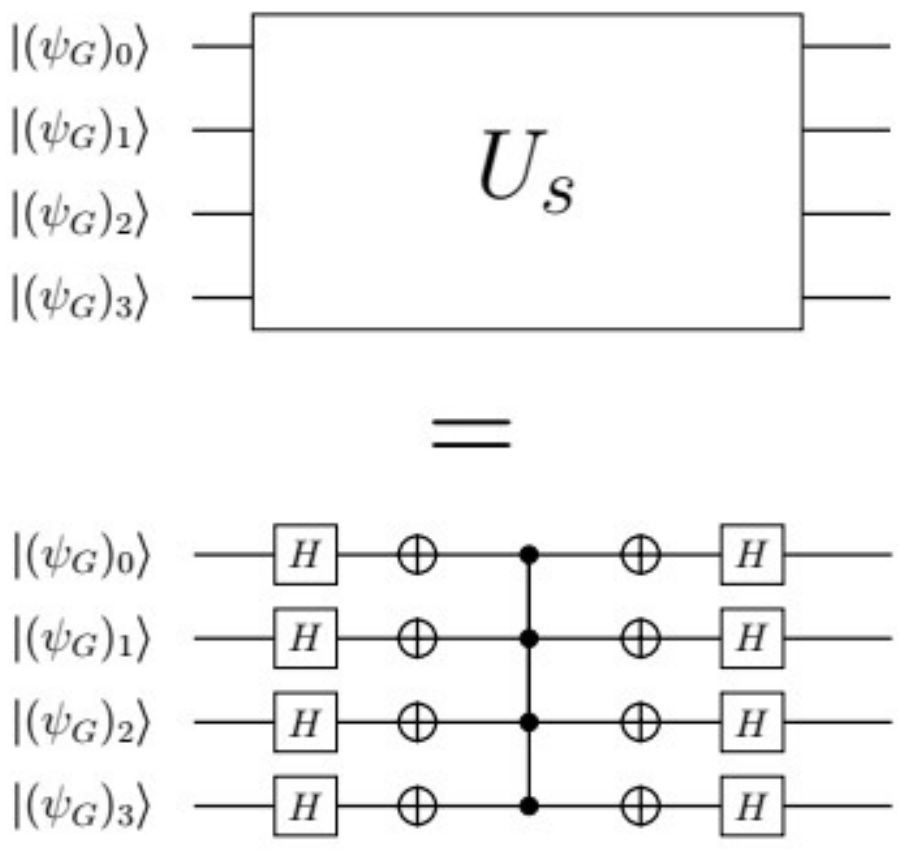}
\caption{Circuit diagram for $U_s$ for $|G|=2^4$.}\label{us}
\end{figure}

The simplest part of Grov to implement is the standard $U_s$ operator as defined in Eq.~\eqref{eq:ref}. As discussed in Ref. \cite{Boyer1998}, if $|G|=2^n$ for some $n$, then $V=H^{\otimes n}$ is given by the Hadamard gate acting on every qubit of the group register. The remaining reflection is implemented by a controlled $\pi$ phase gate on a computational $0$ input, i.e. apply NOT to all qubits  and then apply the multi-controlled $Z$ gate. Finally, we uncompute everything but the multi-controlled $Z$ gate.  An example of this circuit is shown in Fig. \ref{us}. If $|G|$ is not a power of $2$, we only have to modify the change of basis given by $V$ to some other change of basis operator such as the quantum Fourier transform (QFT). The cost of the former is $\mc C(U_s) \sim\mc O(\log |G|)$ while the  latter case scales as $\mc C(U_s) \sim \mc O(\log^2 |G|)$ if we use QFT.

\begin{figure}
\centering
\includegraphics[scale=.5]{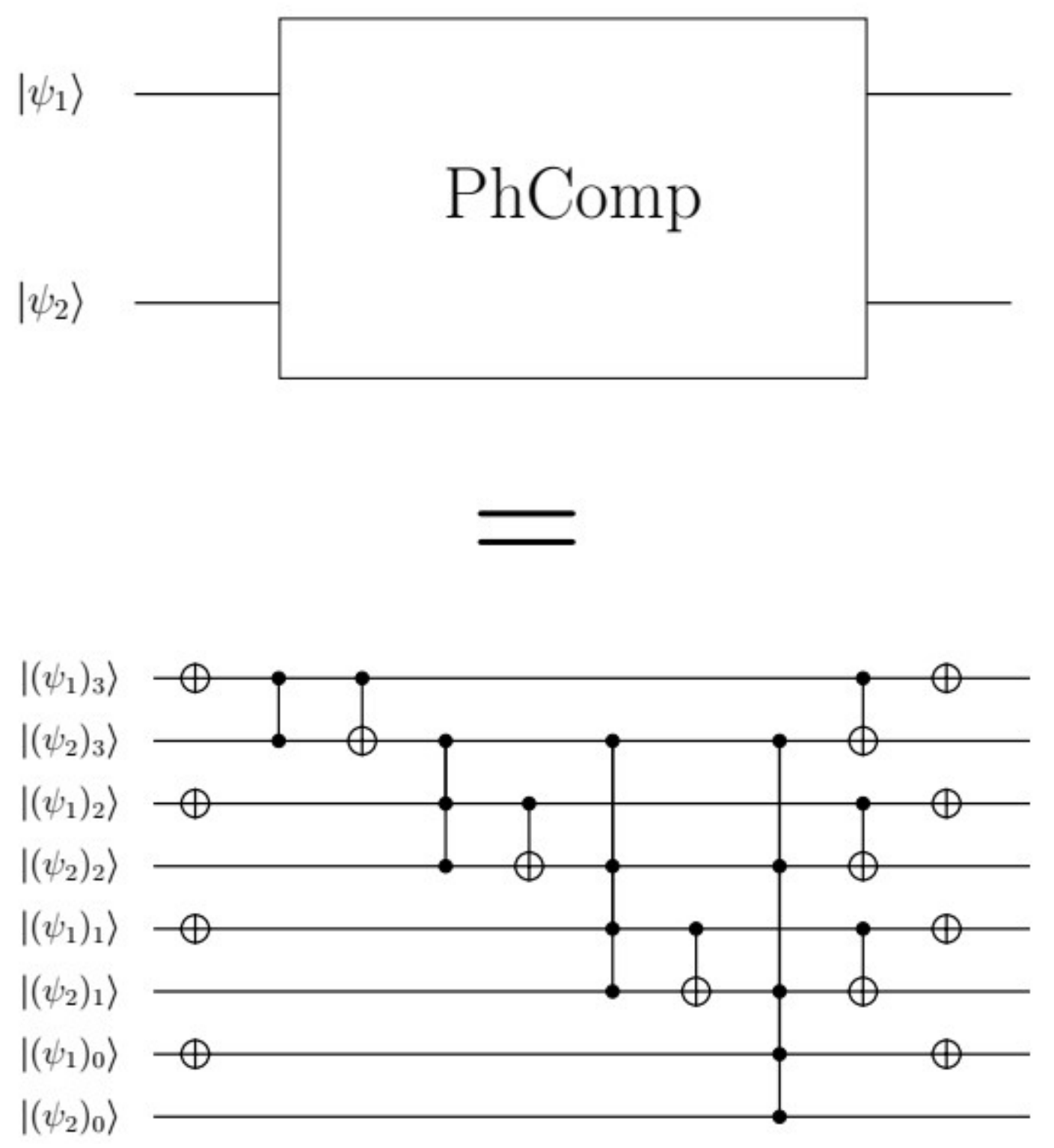}
\caption{Circuit diagram for PhComp for $|V|=2^4$.}\label{phcomp}
\end{figure}

We next consider an implementation of PhComp as defined in Eq.~\eqref{eq:phcomp} by considering a bitwise comparison of the input registers. We start with the most significant bit of the binary expansion of a computational input value and proceed to the least significant. At the $i^{th}$ bit, we need to calculate two binary values, the first representing whether or not we should apply the $\pi$ phase at the current bit, and the second representing whether or not we should continue to compare on the remaining lesser bits. That is, if the two bits differ, the value containing $1$ is greater, so we need to prevent any additional phases from being apply on lesser bits. A truth table for this calculation is given in Table \ref{tbl2} for input bits $a_i$ and $b_i$. From this, we find that $\text{(apply phase)}_i = \overline a_i b_i$ conditioned on the truth (AND-ed with) all greater $\text{(continue)}_j = \overline a_j \oplus b_j$ bits for $j>i$.  So our method of implementing PhComp is to NOT all qubits of the first register $a$ and then compare from the most to least significant qubit. At the $i^{th}$ qubit, we calculate $\text{(continue)}_i$ on $b_i$ using CNOT, but not before calculating $\text{(apply phase)}_i$ in the phase with a multi-control Z gate between $ \overline a_i$, $b_i$ and all the $b_j$ for $j>i$ ( which now contain the (continue) bits). Finally, we uncompute the CNOT and NOT gates. An example circuit is shown in Fig \ref{phcomp}. Assuming the cost of a multi-control $Z$ gate scales linearly with the number of controls, the cost of PhComp is $\mc C(\text{PhComp}) \sim \mc O(\log^2 |V|)$. However, if we have additional ancilla qubits available, we can use these to reduce $\mc C(\text{PhComp}) \sim \mc O (\log |V|)$. See Appendix~\ref{redux} for details.  

\begin{table}[t]
\begin{ruledtabular}
\begin{tabular}{ c c c c }
$a_i$ &$b_i$ & $\text{(continue)}_i$ & $\text{(apply phase)}_i$ \\
\hline
0 & 0 & 1 &0\\
0 & 1 & 0 & 1\\
1 & 0 & 0 & 0 \\
1 & 1 & 1 & 0
\end{tabular}
\end{ruledtabular}
\caption{Truth table used to form PhComp}\label{tbl2}
\end{table}

\begin{figure}
\centering
\includegraphics[scale=.7]{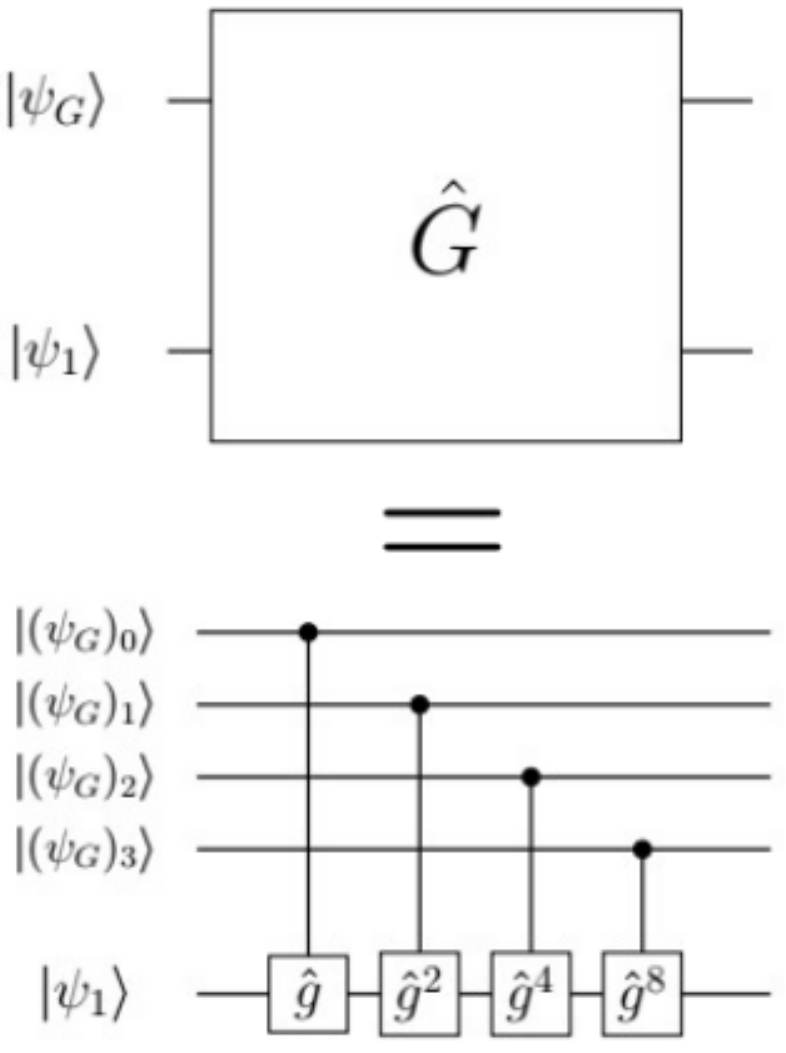}
\caption{Circuit diagram for the group action operator $\hat G$ for a single-cycle abelian group generated by $g$ and simulated by $\hat g$ for $|V|=2^4$.}\label{gact}
\end{figure}

\begin{figure}
\centering
\includegraphics[scale=.4]{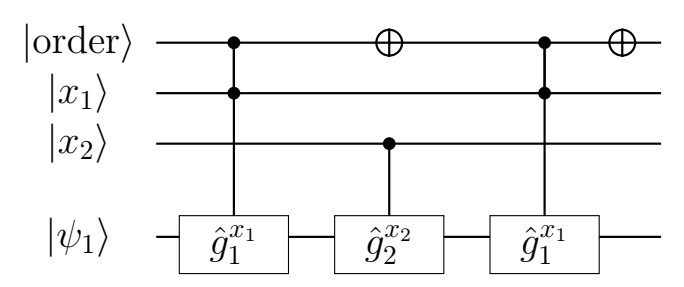}
\caption{Method for implementing a group with two non-commuting generators. $\ket{\text{order}}$ is a single qubit  which determines the order the generators are applied. }\label{nag}
\end{figure}

The form of the group action operator is entirely dependent on the group. We take the simplest case first which is an abelian group with a single cycle, whereby $g(x)= g^x$ for the group generator $g$. We assume we can form a circuit for the operator $\hat g$ acting on a position register which achieves
\begin{align}
\hat g \ket{v} = \ket{g v}.
\end{align}
 We then control $\hat g^{2^i}$ on the $i^{th}$ qubit of the group register as show in Fig. \ref{gact}. This method can then be generalized to multi-cycle abelian groups by subdividing the group register so there is one subregister for each cycle and generate a circuit similar to Fig. \ref{gact} for each cycle. If the group is non-abelian, one has to consider a strategy for indexing powers of the generators and their order. For example, suppose the group is generated by two non-commuting operators $g_1$ and $g_2$. Each generator forms its own abelian subgroup so we can use the same strategy for them separately and with their own sub-group register. Furthermore, the order for applying these operators can be controlled by a single qubit, $\ket {\text{order}}$, using the circuit in Fig. \ref{nag}. If $\ket{\text{order}} = \ket{0}$, then the group operator applied is $g_1^{x_1}g_2^{x_2}$ and if $\ket{\text{order}}=\ket{1}$ then the group operator applied is $g_2^{x_2} g_1^{x_1}$. We note this may not be the most efficient method in terms of qubit use for the group register qubits. For example, if $x_2=0$, then the state of $\ket{\text{order}}$ doesn't matter and so there are redundant index states in the group register. This is also the case if there are redundancies in the order of non-zero powers of the generators, i.e. $g_1^{x_1} g_2^{x_2}=  g_2^{x^\prime_2}g_1^{x^\prime_1}$, for some values of the indices. The most efficient method depends on the group, but we include this example to demonstrate that, in principle, one can handle non-abelian groups using roughly the same strategy as was used for abelian groups. 

The scaling of $\hat G$ is highly dependent on the group being used, but it should be clear that in many reasonable cases, the scaling should be $\mc C(\hat G) \sim \log(|G|)\mc C( \hat g)$ where we assume the generators can be implement at cost $\mc C( \hat g)\sim \mc C(\hat g^n) \sim \text{polylog}(|V|$) for any power $n$. That is, implementing the power of a generator must not scale with that power. To demonstrate the importance of this, consider the single-cycle abelian case. If we implement $\hat g^2$ with two copy of $\hat g$ and so on for the other powers, then 
\begin{align}
\mc C(\hat G) \sim \mc C(\hat g) \sum_{k=1}^{\log |G|} 2^k \sim \mc C(\hat g) |G|,
\end{align}
Clearly, this is not efficient, and our oracle scales with the size of the search space. However, if the implementation of the powers of $\hat{g}$ can be simplified so as to scale on the order of $\hat g$ or less, then we achieve our desired scaling $\mc C(\hat G) \sim \mc C(\hat g) \log |G|$.  It is reasonable to believe this is possible in the general case. Suppose we take for granted the complexity of a quantum circuit corresponding to a periodic operator scales with the size of its period. $\hat g^2$ has half the period of $\hat g$ and $\hat g^4$ has half the period of $\hat g^2$ and so on. So one would expect that  $\hat g$ is actually the most expensive power to implement. %This does, however, provide a challenge for current circuit simplification schemes, which rely on %pushing \noteSJ{Is there a better word here than 'pushing'?}
%exchanging the order of commutative gates to cancel and local pattern matching \cite{Maslov2003}.

\begin{figure}
\centering
\includegraphics[scale=.45]{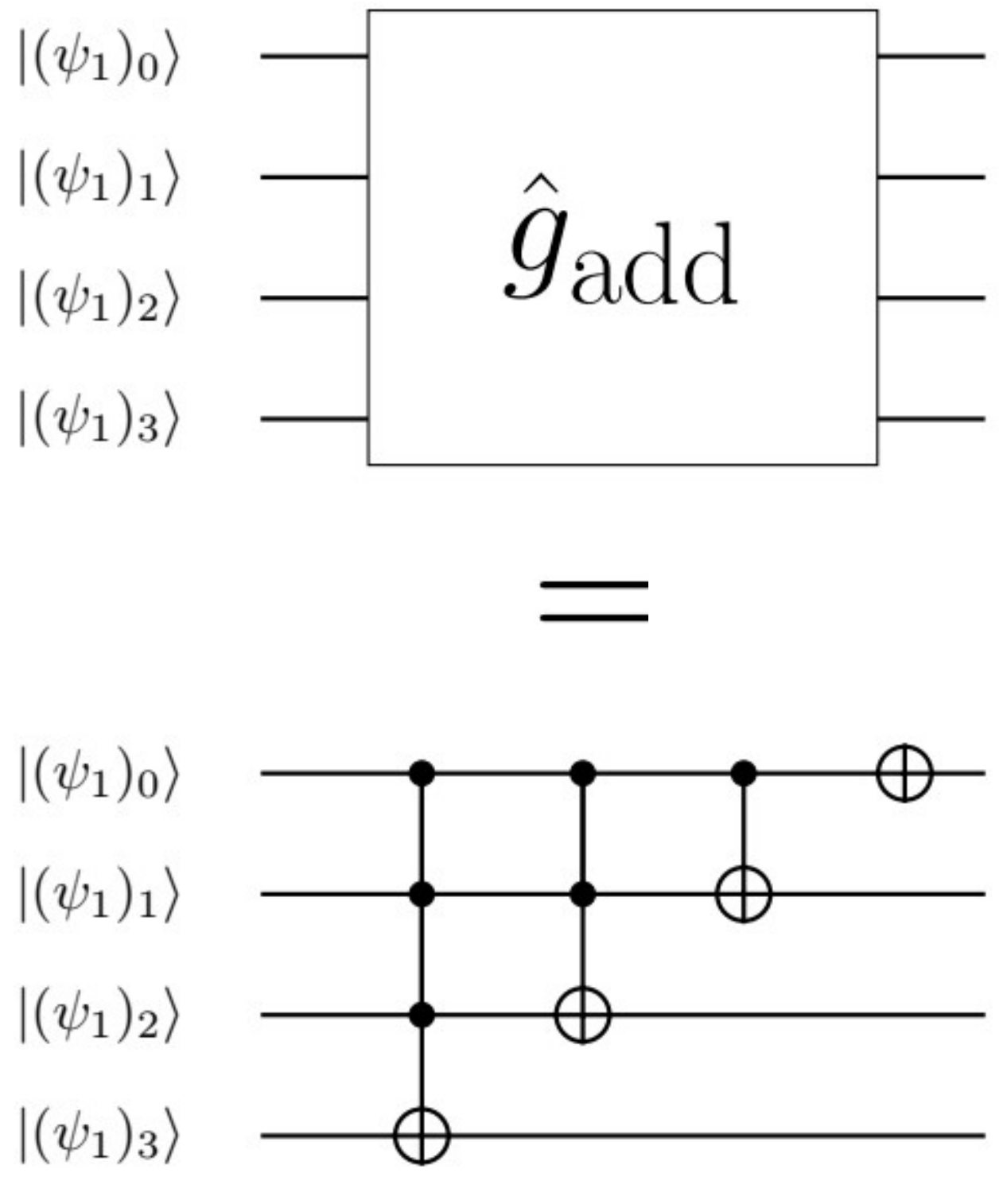}
\caption{Circuit diagram for $\hat g_{\text{add}}$ for $|V|=2^4$.}\label{add}
\end{figure}

 To make this discussion more concrete, consider the  example of the group representing addition mod $N=2^n$ for some $n$ which we denote $G_{\text{add}}^{N}$, i.e.
\begin{align}
\hat G_{\text{add}}^{N} \ket{x}\ket{y} = \ket{x}\ket{ x+y},
\end{align}
%\noteSJ{Is this correct?}AS: Yes?
%
where mod $N$ is implicit. Implementing this operator using the methods discussed here\footnote{We are aware of better in-place adders, namely those which calculate the addition in the phase or via some ripple-carry scheme\cite{Cuccaro2004}. We stick to this less efficient implementation of the adder as it imitates our more generic construction of $\hat G$.}, one can use $\hat g_{\text{add}}$ consisting of a sequence of multi-control NOT gates as shown in Fig. \ref{add} (where we recall that $\hat g_{\text{add}} \ket{y}= \ket{y+1}$). It is clear that $\hat g_{\text{add}}^2$ is given by removing all gate action and control lines on the least significant bit, and so on for the other powers. For such a simple case, it's easy to see this simplification, but for a compiler which only moves commutative gates and considers local pattern matching, this dramatic reduction might go unexploited, so a manually-optimized implementation might be preferred.

%As a counter example that we can not always see this reduction, suppose our Hamitonian is some model of a spin chain with translation symmetry. Naturally, each qubit of the position register would correspond to one spin of the material system and so $g_{\text{spin}}$ would consist of a cascade of nearest-neighbor swaps, $g_{\text{spin}}^2$ next-nearest-neighbor swaps and so on. So if our quantum computer has full connectivity between all its qubits, we see the reduction for larger powers of the generator, but if the position register itself is limited to nearest-neighbor connectivity, then we have not choice but to just double the number of swaps.  It worth note that $g_{\text{spin}}$ scales as $\mc O (|G|)$ or greater no matter what, but this is because the group is already exponentially smaller than the position space. It may still be advantageous to use a quantum computer in this case since applying the group action is so natural for the quantum computer, even though we have $\mc O\left(|G|^{\frac{3}{2}}\right)$ or greater scaling for the algorithm runtime.

It is worth considering the specific case of $G$ for spin Hamiltonians as this is the most natural use for a quantum solution to this problem. The natural mapping of the problem would assign one qubit to each spin in the physical system, and most geometric symmetry generators such as those for translation or rotation (as opposed to spin symmetries), can be simulated by a $\hat g_{\text{spin}}$ consisting of swap gates. For example, translation on a spin chain would use a $\hat g_{\text{spin}}$ consisting of a cascade of nearest-neighbor swaps. Note that here and in general,  $ \mc C(\hat g_{\text{spin}}) \sim \mc O(|G_{\text{spin}}|)= \mc O(\log |V|)$, but this is because the group is already exponentially small in the size of $V$. So in general we expect $\mc C(\hat G_{\text{spin}}) \sim |G_{\text{spin}}| \log|G_{\text{spin}}|$.  

In terms of comparing costs with the classical on-the-fly method, we expect $C(G)$ to be of the order of $\mc C(\hat G)$ in which case the quantum solution out performs the classical one. The classical divide \& conquer method has a smaller constant coefficient, so the quantum solution outperforms it for relatively larger group sizes, but always uses exponentially less memory.

\subsection{Oracle Budget and Probability of Success}\label{sec:obandps}

\begin{figure}
\centering
\includegraphics[scale=.55]{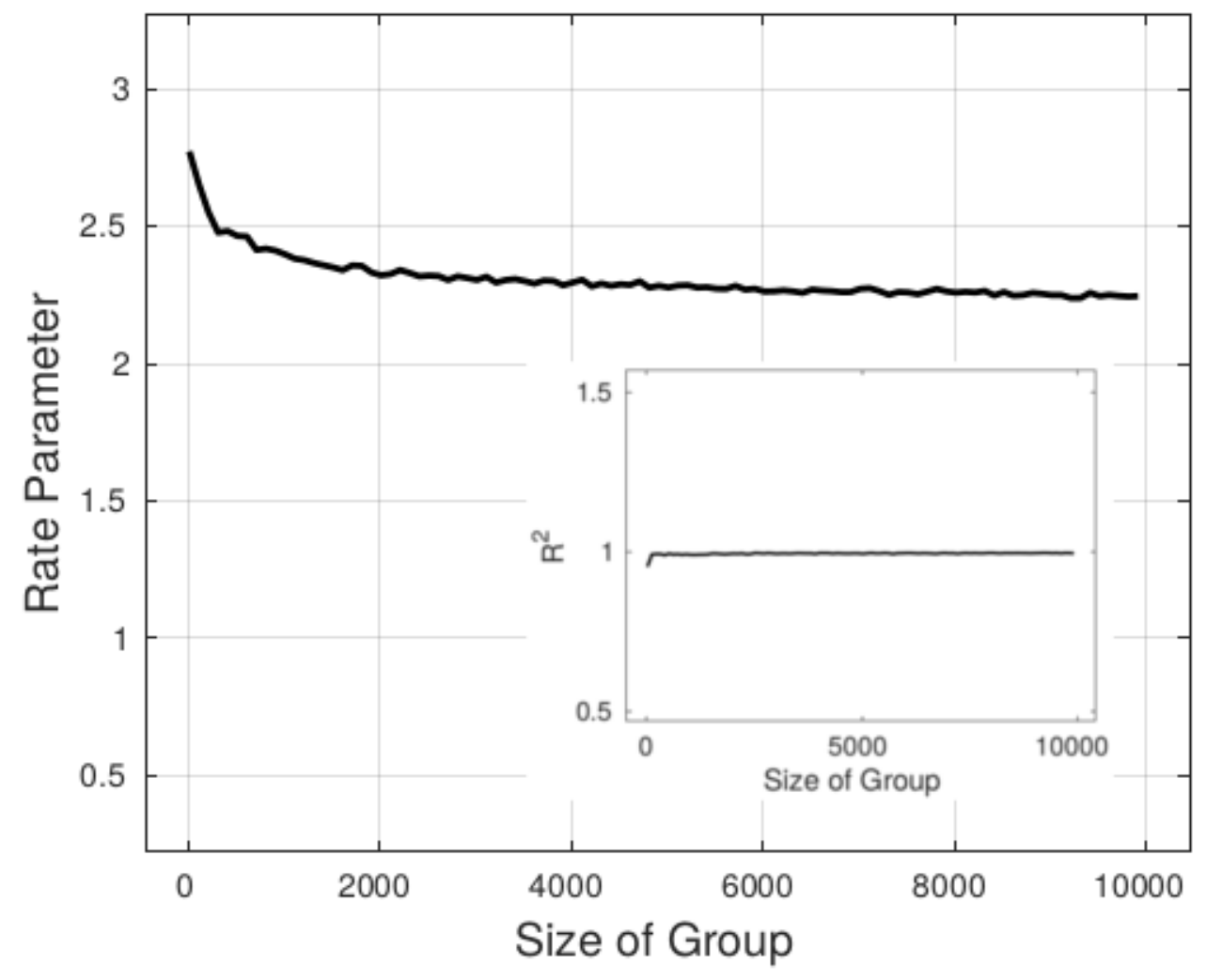}
\caption{Plot of the rate parameter as a function of group size from the simulation of the classical parts of Algorithm~\protect \ref{alg1}. The insert shows the $R^2$-value of the linear regression used to derive the rate parameter; $\beta=0.95$, $\gamma =1.15$.}\label{fig:em1}
\end{figure}

\begin{figure}
\centering
\includegraphics[scale=.5]{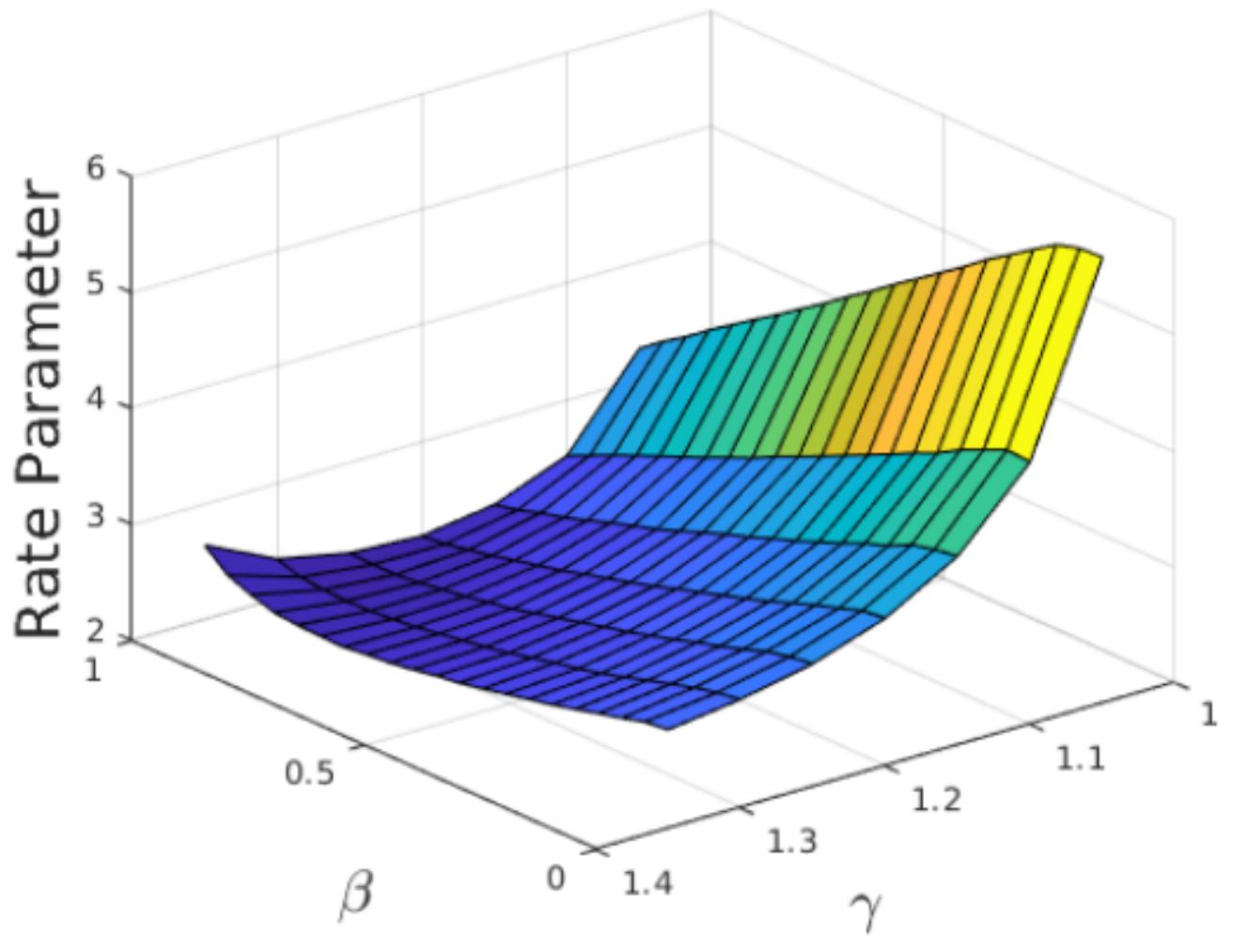}
\caption{Plot of the rate parameter $a$ as a function of $\beta, \gamma$.}\label{fig:search}
\end{figure}

To complete the algorithm, we need to determine the constants $\alpha, \beta$ and $\gamma$. As we have the exact solution for the probability of success for a single Grover search \cite{Grover1996, Boyer1998} (line 9-11 of Algorithm~\ref{alg1}), we are able to simulate the classical parts of the algorithm using $G_{\text{add}}^{N}$ as the group in order to determine the behavior of these parameters. Note that, without error, the oracle query complexity is unaffected by the details of the group (aside from its size), so the following results should be general. As we know the solution to the orbit representative problem for this trivial example, we can run the simulation until the correct answer is obtained. This allows us to empirically determine the probability of success as a function of the total number of calls. For a window of probabilities $P_{\text{success}} \in [0.2, 0.995]$\footnote{We clearly don't care about probabilities less that $20\%$ and beyond $99.5\%$, the rate of increase of the probability is hard to discriminate within our simulation.}, we find that the asymptotic form of the probability for large $N$ is given by
\begin{align}\label{eq:prob}
P_{\text{success}} \sim 1- \exp\left(-\frac{T^2}{a^2 N}\right),
\end{align}
where $T$ is the number of oracle calls and $a$ is the {\it rate parameter} which is a function of only $\beta, \gamma$ and is empirically determined. By linearizing Eq.\eqref{eq:prob} with $\frac{1}{a}$ as the slope, we can calculate the rate parameter as is the case in Fig.~\ref{fig:em1} as well as demonstrate this is the correct asymptotic form. One can see that the $R^2$-value of the linear regression asymptotically approaches $1$ and the rate parameter approaches a constant for fixed $\beta, \gamma$. This allows us to determine the oracle budget parameter $\alpha$. For a given application, if we allow for a probable error in the solution of $\epsilon>0$, then $\alpha$ is given by 
\begin{align}
\alpha \sim a \sqrt{ -\ln \epsilon}.
\end{align}
So we want to determine the values of $\beta$ and $\gamma$ such that we minimize $a$.  Figure \ref{fig:search} shows a survey of $a$ as a function of $\beta$ and $\gamma$. From this, we have chosen $\gamma= 1.15$ and $\beta =0.95$ as the near optimal values. This value of $\gamma$ is near previously discussed values, where Ref.~\cite{Boyer1998} suggests $\frac{6}{5}$. However, $\beta$ being near one suggests we gain a good deal of information knowing that the number of marked items has decreased from one call of Gsun to another. For comparison, if we use $\epsilon=0.5$ and $a\approx 2 \hyph4$, the resulting oracle budget parameter is $\alpha \approx 1.6\hyph 3.3$ which is a considerable reduction compared to $\alpha \approx 5.6$ for the analytic value found in Appendix~\ref{derive}. For applications which require a high probability of success i.e. $\epsilon=0.01$, we obtain $\alpha \sim 4.3 \hyph 8.6$.

\section{Full Simulation for a Perfect Quantum Machine}

To check the behavior of Algorithm~\ref{alg1}, we implement a full quantum simulation using the Intel Quantum Simulator (Intel-QS)\cite{qHIP} and $G_{\text{add}}^{2^n}$ as our group for $n=4\hyph 8$. This requires $12\hyph 24$ qubits using no additional ancilla to reduce the depth of the quantum circuits. Although $G_{\text{add}}^{2^n}$ is not a useful problem instance, it does maximize the group size relative to the number of positions, i.e. $\log|G|=\log|V|$ and so this represents the most efficient benchmark using the fewest qubits. Furthermore just as with the purely classical simulation from the last section, knowing the correct answer allows us to avoid choosing an oracle budget, and instead run the algorithm until the solution is found to determine the probability of success as a function of total number of calls\footnote{ Still, we do choose a hard stop of $\alpha=\frac{45}{2}$ which as we established is on the high side for a reasonable oracle budget.}. As our simulation is exact, i.e. we are treating the quantum machine as perfect, the details of the group do not affect the results. All quantum subroutines are implemented according to the discussion in Section~\ref{sec:circ}, where multi-controlled gates have been broken down to one- and two-qubit gates using methods from Ref.~\cite{Barenco1995}. This was done to better simulate the algorithm acting on real hardware once noise is added in Section~\ref{sec:simerror}.

\begin{figure}
\centering
\includegraphics[scale=.55]{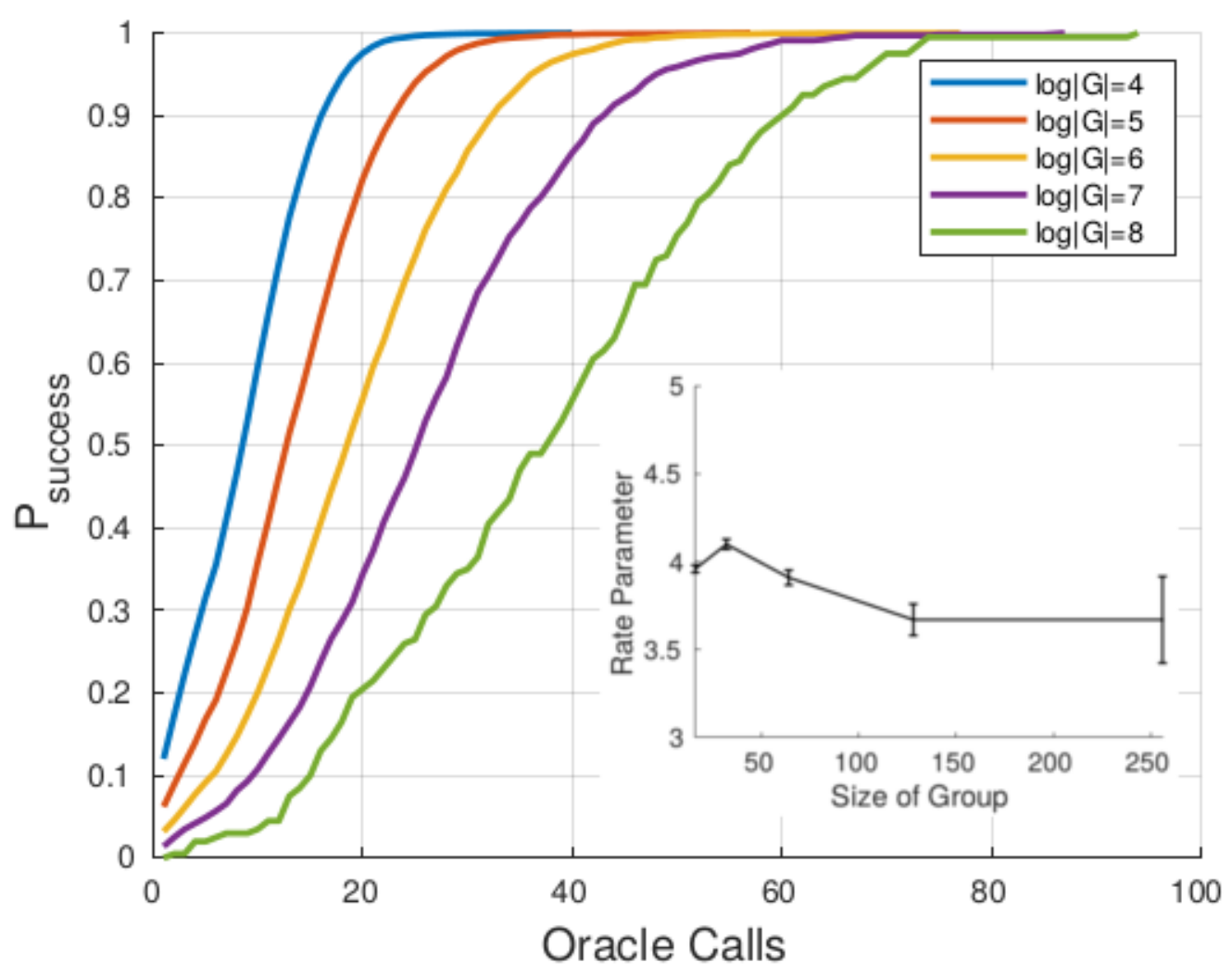}
\caption{Plot of the simulated value of $P_{\text{success}}$ as a function of number of oracle calls for various group sizes.  The insert shows the rate parameter for these different group sizes. The number of trials used is 10000, 10000, 4000, 1000 and 200, respectively. }\label{fig:clean}
\end{figure}

Figure~\ref{fig:clean} shows the probability of success as a function of oracle calls, where the insert shows an effective rate parameter. We note that the probability in Eq.~\eqref{eq:prob} is asymptotically correct in the limit of large $N$ and as such, the curves for these smaller group sizes do not fit this form well. Instead we define the effective rate parameter, $a_{\text{eff}}$ via
\begin{align}\label{eq:aeff}
    \frac{1}{a_{\text{eff}}\sqrt{N}} = \text{avg}\left(\text{diff}_T \left(\sqrt{-\ln\left(1-P_{\text{success}}\right)}\right)\right)
\end{align}
where we treat $P_{\text{success}}$ as a function of oracle calls, $T$, and $\text{diff}_T$ is the difference between two successive values of $T$. Despite the poor fit, $a_{\text{eff}}$ is still indicative of the trends. We then determine error bars for $a_{\text{eff}}$ via
\begin{align}
\delta a_{\text{eff}}= \sqrt{\frac{N}{M}}\sigma_{\text{eff}} a_{\text{eff}}^2,
\end{align}
%
%\noteSJ{$\sqrt{N}$ should be in the numerator. If so, also update error bars in the figure.} AS: error bars are correct. I just messed up the formula in the text.
where $\sigma_{\text{eff}}$ is the standard deviation of the expression which is averaged in Eq.~\eqref{eq:aeff} and $M$ is the number of trials. From Fig.~\ref{fig:clean}, we find the behavior as expected from the classical simulations. We notice however that the effective rate parameter is higher for the full simulation as compared to the classical simulation. Although not optimal, the effective rate parameter still suffices. For example, if we desired a $99\%$ chance of success and we chose the rate parameter to be $a= 4$ ($\alpha \approx 5.7$), then the oracle budgets would be $23, 32, 45, 64$ and $91$, respectively, for the group sizes shown in Fig.~\ref{fig:clean}. From the figure, we see that we would achieve nearly or better than our target $99\%$ chance of success.

\section{Error Mitigation Strategies and their Simulation}

One of the benefits of Gmin is that the best-known value for the minimum is alway monotonically decreasing with the number of oracle calls. Unlike Grover search, this is true even for a faulty implementation on an imperfect quantum machine. Furthermore, vagaries of faulty implementation are partially compensated for by the classical random sampling of the number of oracle calls for any single coherent Grover search. Put a different way, though we allot a set oracle budget, not all these calls are implemented in a single coherent step.  This suggests Grover minimization is a reasonable use for near-term, noisy hardware. Still, noise has its costs. In this section, we describe some strategies for mitigating the cost of errors. We then simulate some of these methods to determine their effectiveness.

\subsection{Strategies}

We start by describing two error mitigation strategies. As mentioned, the approach to a solution is monotonic regardless of the error rates. Thus the most obvious method is to simply increase the oracle budget, leaving all else the same, a method we refer to as {\it static} error mitigation (SEM). The obvious downside to this method is that the increase in the oracle budget would reasonable need to scale with the size of the system --assuming roughly independent error rates for each qubit--in which case, we may lose our quantum advantage. This is supported by analytic results on Grover search with a faulty oracle in Refs.~\cite{Shenvi2003, Regev2012}, where for certain toy error models, the polynomial quantum speed-up is either partially or entirely lost.  

The other strategy takes advantage of the additional qubits which do not hold the search space. The two position registers are included only as a means of marking elements of the search space and implementing the oracle. As such, they should hold the same computational basis value at the beginning and end of a single call to Grov. This allows us to measure these registers without disturbing the coherence of the group register which is responsible for the quantum speed-up. Moreover, any terms in the full state of the system (as expanded in the computational basis) which hold values in the position registers which differ from $v$ and $v_{\text{best}}$ are in error and measuring the correct values projects the system back to an un-errored, or at least less-errored state. Thus we suggest the following: at the end of any call to Grov, measure the two position registers. If their measured values differ from that of the classically stored values $v$ and $v_{\text{best}}$, we abort the remaining Grover steps on line 10 of Algorithm~\ref{alg1} and go back to step 8, for which the errored oracle calls do not count against our oracle budget. It is important to note that we do not randomly sample $p$ again as this would introduce a bias toward smaller values of $p$ as they are less likely to experience an error. We refer to this strategy as {\it active} error mitigation (AEM). This is because the total number of oracle calls, both errored and un-errored, is not fixed, but depends on the rate of error.

The downside of this method is that all the oracle calls up to the point an error is found still cost time which is now wasted due to the error state. To mitigate this waste, before restarting the Grover search, we measure the group register and continue to check to see if a better value is found. To do so is practically free (up to one additional effective oracle call to perform the check) and it can only increase our chances of finding the minimum, even if by a minuscule amount. Moreover, simulations demonstrate the increase is significant. We refer to this as a {\it measure-and-check} strategy.

All together, the AEM version of the Gmin algorithm is presented in Algorithm 2. Note we have added a hard stop for total number of oracle calls as characterized by $\ell$ to avoid infinite run-time. Ideally AEM ``protects'' the probability of success for a fixed oracle budget and a large range of error rates. That is, $P_\text{success}$ as  a function of un-errored oracle calls (i.e. as a function of the $c_1$ count in Algorithm~\ref{alg2}) takes the form of Eq.~\eqref{eq:prob} with a rate parameter which is only weakly dependent on the error rates. Again, the downside is the non-deterministic run-time which can bloat if the error rate is too high.

\begin{algorithm}[H]
\caption{AEM Grover Minimization} \label{alg2}
\begin{algorithmic}[1]
\State Allocate QRegister $\ket{\psi_G}$ of size $m$
\State Allocate QRegister $\ket{\psi_1}$ of size $n$
\State Allocate QRegister $\ket{\psi_2}$ of size $n$
\State $v_{\text{best}}=\gets v, x_{\text{best}}\gets 0, \text{good} \gets \text{true}$
\State $c_1 \gets 0, c_2 \gets 0, t \gets 1$
 \While{ $c_1< \alpha \sqrt{|G|}$ AND $c_2< \ell |G|$}
\If {good} 
\State $p\gets $rand$(0,t-1)$\;
\Else 
\State good $\gets$ true
\EndIf
\State Initialize$\left( \ket{\psi_G} \otimes\ket{\psi_1} \otimes\ket{\psi_2} \gets \ket{0}\otimes\ket{v} \otimes \ket{v_\text{best}}\right)$\;
 \State $V \ket{\psi_G}$\;
\For{$i \in [1:p]$}
\State$\text{(Grov)} \ket{\psi_G}\ket{\psi_1}\ket{\psi_2}$ \;
\State Measure$(v_1 \gets \ket{\psi_1}, v_2 \gets \ket{\psi_2})$
\If{$v_1 \neq v$ OR $v_2 \neq v_{\text{best}}$}
\State good $\gets$ false
\State $c_2 \gets c_2 +i +1$
\State break {\bf for}
\EndIf
\EndFor
\State Measure$(x \gets \ket{\psi_G})$\;
 \If{good}
\State $c_1 \gets c_1+p+1$
\State $c_2 \gets c_2+p+1$
\EndIf
  \If{ $f_v(x)< v_{\text{best}}$}
 \State $\tilde v_{\text{best}}\gets f_v(x)$\;
\State $ x_{\text{best}} \gets x$\;
 \State $ t\gets \max(1, \beta t)$\;
\Else
\If {good} 
\State $t\gets \min (\gamma t, \sqrt{N})$\;
\EndIf
\EndIf
 \EndWhile
\State \Return $v_{\text{best}}, x_{\text{best}} $
\end{algorithmic}
\end{algorithm}

\subsection{Performance of AEM Gmin}\label{sec:scale}

In Appendix~\ref{ap:AEMper}, we analyze the performance of AEM Gmin using a simple error model. Let the average qubit lifetime $\braket{t}$ (say the average between $T_1$ and $T_2$ as described below) scale as
\begin{align}
\braket{t} \sim\frac{\delta}{4}\mc C(\text{Grov})\sqrt{N},
\end{align}
for some $\delta>0$. Then we find that optimally (such that $e=1$; see Appendix~\ref{ap:AEMper} for details) the probability of success for $p$ AEM Grov calls, including measure-and-check when an error is found, is asymptotically 
\begin{align}\label{eq:aemprob}
P&_{\text{success}}^{(p)}\nonumber \\
&\sim \frac{\delta^2}{1+ \delta^2}\left( \frac{(1-\sigma^p)}{2} + \sigma^p \sin^2\left((2p+1)\theta\right)\right),
\end{align}
where $\sigma\sim \exp\left(-\frac{\mc C(\text{Grov})}{\braket{t}}\right) =\exp\left(-\frac{4}{\sqrt{N} \delta}\right)$ is the probability of having no error in a single call to Grov. Recall that the probability of success in the absence of noise ($\delta\to \infty$) is $\sin^2\left((2p+1)\theta\right)$. Without measure-and-check after the error, this probability is degraded to $\sigma^p \sin^2\left((2p+1)\theta\right)$, in which case the probability of success is exponentially sensitive to the value of $\delta$. With measure-and-check, the probability is only polynomially sensitive to $\delta$. 
%From this expression, we can see that as the clean Grover search term, $\sin^2\left((2p+1)\theta\right)$, is degraded by the factor of $s^p$, the AEM term, $(1-s^p)$, partially compensates. Moreover, without AEM, the probability of success is exponentially sensitive to the value of $\delta$ whereas with AEM, the probability is only algebraically sensitive to the value of $\delta$. 
To demonstrate this, consider the case when we are searching for a single element, $p\sim \sqrt{N}$ and so $\sin^2((2p+1)\theta)\sim 1$. If $\delta= 4$, then the AEM probability of success with measure-and-check goes as $\frac{16}{17}\left(\frac{1-\exp(-1)}{2} + \exp(-1)\right) \approx 64\%$, which is reasonably better than the no-measure-and-check probability of $\exp(-1) \approx 37 \%$. However,  if $\delta=1$, then the AEM probability is $\frac{1}{2}\left(\frac{1-\exp(-4)}{2} + \exp(-4)\right) \approx 25 \%$ as compared to $e(-4) \approx 2 \%$. If we go even further and take $\delta =\frac{1}{2}$, the AEM probability of success goes as $\frac{1}{10}= 10 \%$ whereas without measure-and-check, it is negligible. So even though we need the coherence time to scale as $\sim\sqrt{N}$, AEM Gmin is far more forgiving for a smaller value of the coefficient $\delta$. This is further demonstrated numerically in Section~\ref{sec:realsim}. 

The analysis given in the appendix is a general result for Grover search with the same measure-and-check AEM strategy. For AEM Gmin, the fact that the probability of success of a single search is necessarily degraded by noise means we still need to increase the oracle budget in order that the target overall probability of success is maintained. This is done automatically by not counting errored oracle counts.

\subsection{Simulation of Error Mitigation Strategies }\label{sec:simerror}

\begin{figure*}[t]
\centering
\begin{tabular}{cc}
\subfloat[rate parameter for $\log|G| =4$\label{fig:compa}]{\includegraphics[scale=.55]{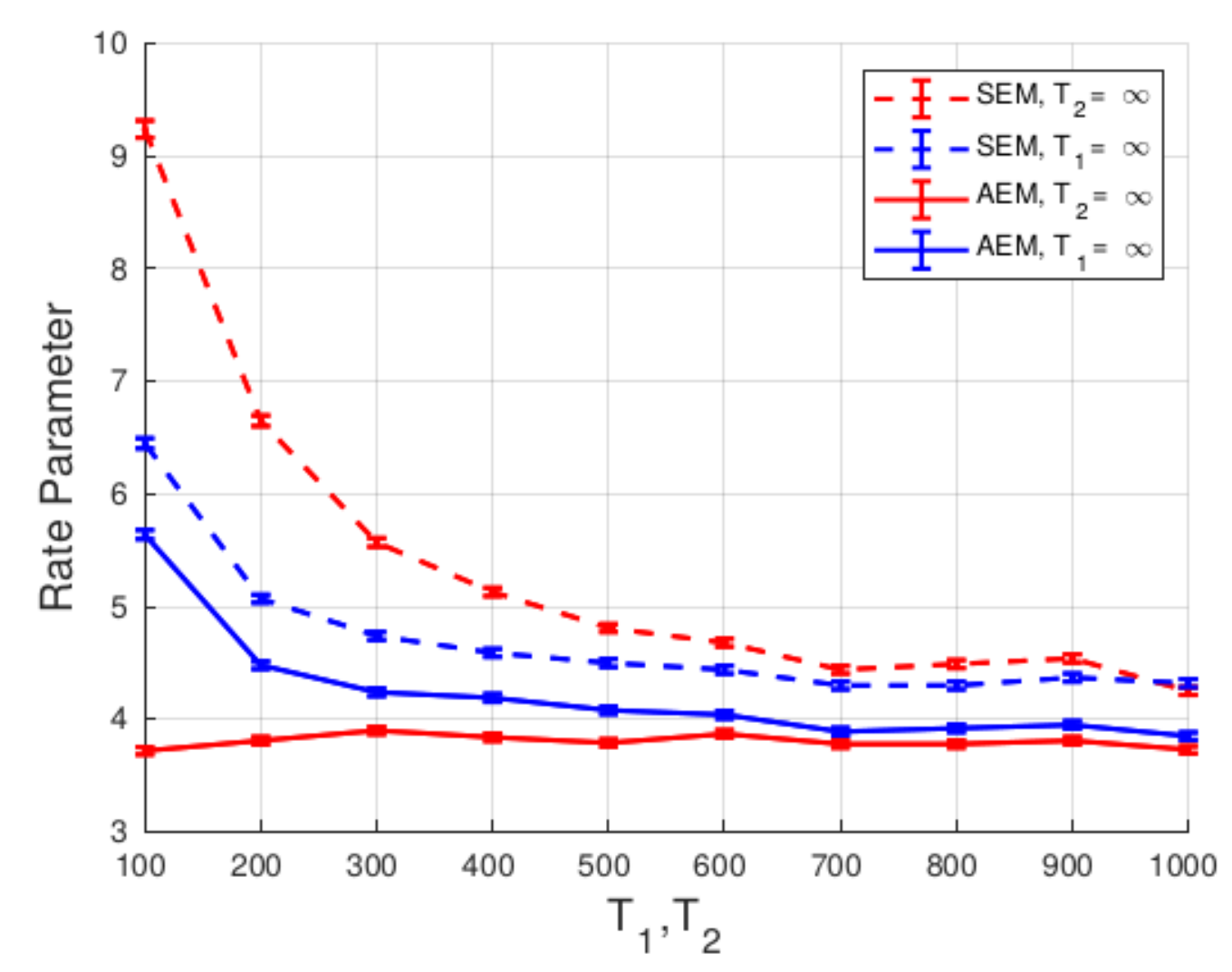}}&
\subfloat[rate parameter for $\log|G| =5$\label{fig:compb}]{\includegraphics[scale=.55]{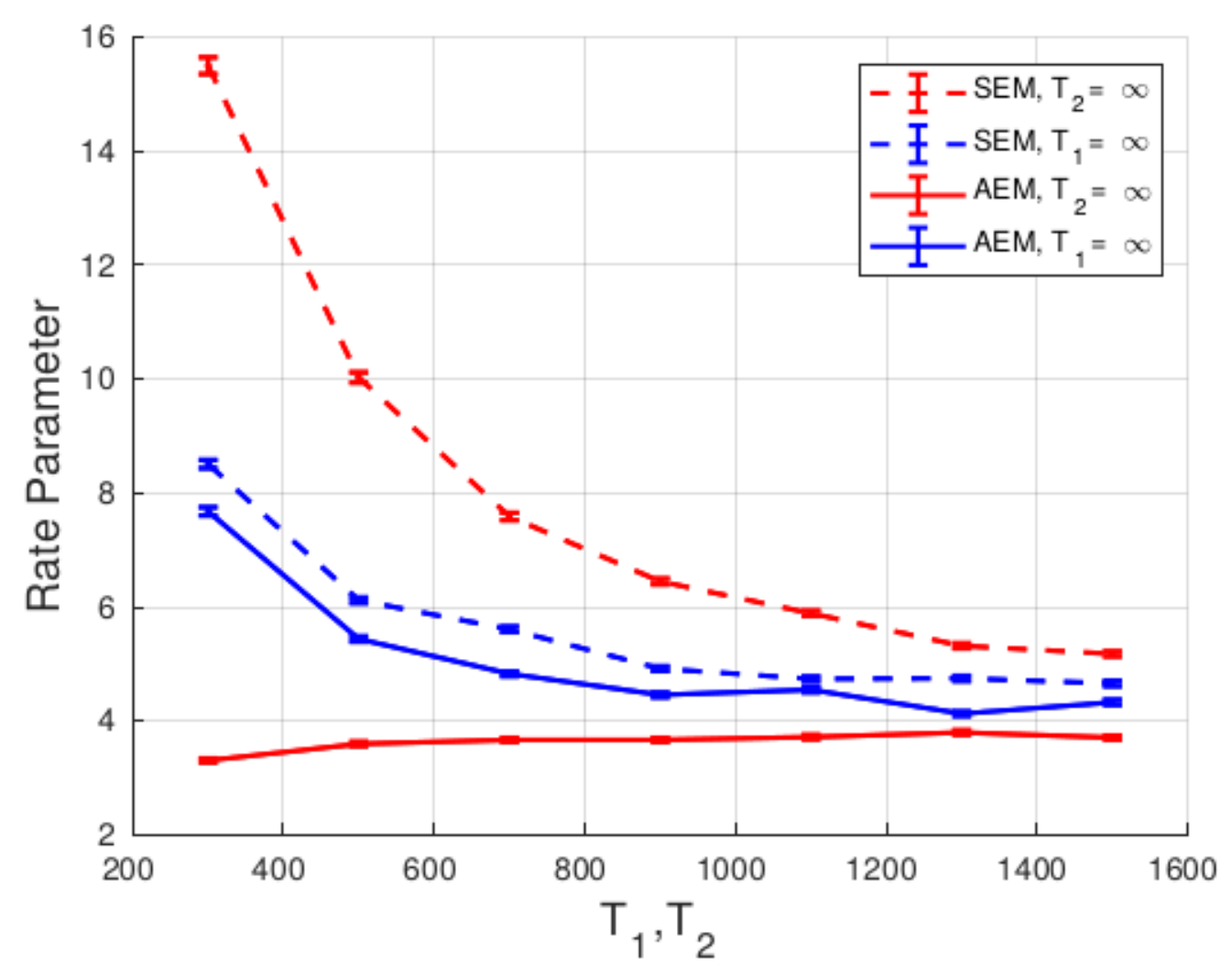}}\\
\subfloat[average run-time for $\log|G| =4$.\label{fig:compc}]{\includegraphics[scale=.55]{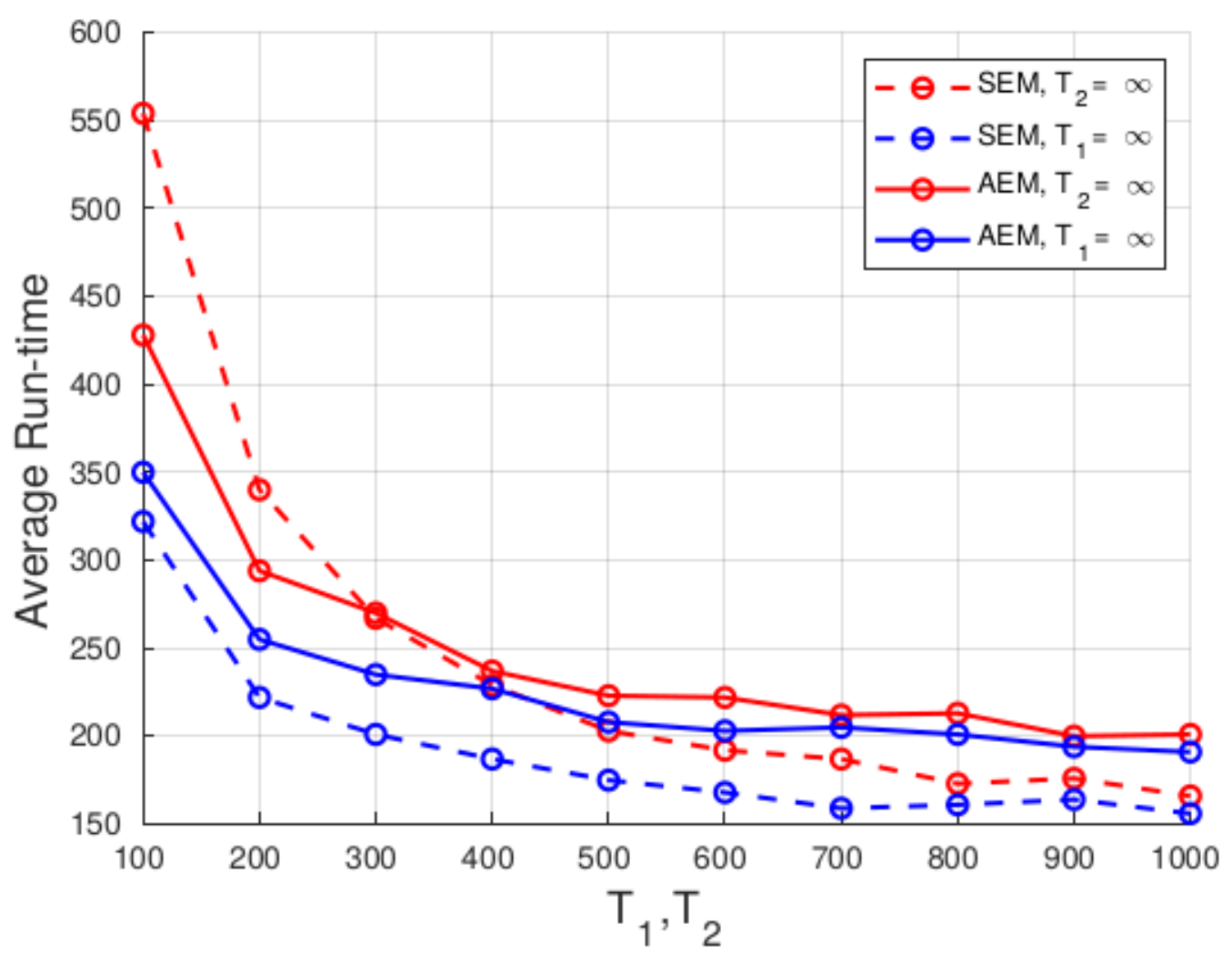}} &
\subfloat[average run-time for $\log|G| =5$.\label{fig:compd}]{\includegraphics[scale=.55]{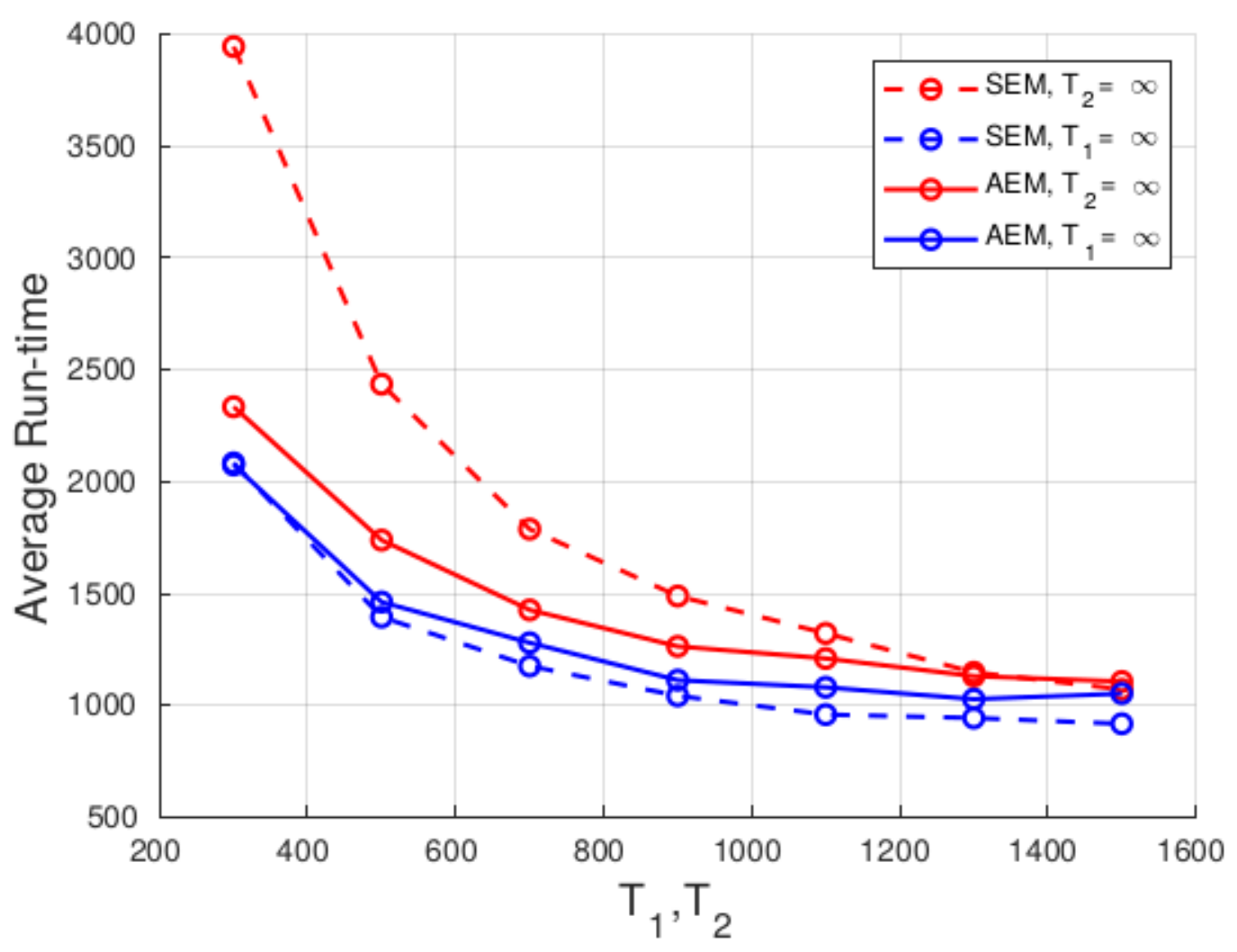}}
\end{tabular}
\caption{Simulation plots for rate parameter and average run-time to contrast SEM (dashed) versus AEM (solid). For each plot, either $T_1$ (blue) or $T_2$(red) are fix at $10^{9} \sim \infty$ and the other is varied. In all cases, the number of trials is  4000. $T_1$, $T_2$ and average run-time are measured in units of the single-qubit gate time.}\label{fig:comp}
\end{figure*}    

To simulate noisy hardware, we used the error model included in the Intel-QS package which is based upon the Pauli-twirling approximation error model \cite{Geller2013}. In this model, before a gate is applied, a random single qubit rotation is applied to each qubit acted on by that gate. The error unitary is given by
\begin{align}
U_{\text{error}} = \exp\left(i v_x X + iv_y Y + iv_z Z\right),
\end{align}
where $X, Y, Z$ are the single-qubit Pauli operators. $ v_x, v_y$ and $v_z$ are parameters chosen at random from a Gaussian distribution whose variance grows with the time from the last gate action in units of the hardware dependent parameters $T_1, T_\phi$ and $T_2$ respectively. %$v_x$ is associated with $T_1$, $v_y$ is associated with $T_\phi$ and $v_z$ is associated with $T_2$.
As $X, Y$ and $Z$ are dependent on one another, the parameters are related by
\begin{align}
\frac{1}{T_\phi} =\frac{1}{ T_2} - \frac{1}{2 T_1}.
\end{align}
Because $T_1$ is associated with the $X$ Pauli operator which flips the computational state, we can think of $T_1$ as the ``bit-flip'' error rate. Likewise, $T_2$ is associated with the $Z$ Pauli operator which applies a $\pi$ phase, so we can think of this as the ``phase-flip'' error rate. To accurately accommodate for this non-deterministic, measurement-based algorithm, some modifications had to be made to the Intel-QS. See Appendix~\ref{detail} for details.

Simulations for both SEM and AEM are shown in Fig.~\ref{fig:comp} for $\log|G| =4,5$ where we have fixed either $T_1$ or $T_2$ to be a large, effectively infinite constant and varied the other. This allows us to determine the effect of each kind of error. $T_1$ and $T_2$ are measured in units of the single-qubit gate time (SQGT); see Appendix~\ref{detail} for details.

In terms of bit-flip error, we can see that AEM does protect the rate parameter over the values of $T_1$ shown in Fig.~\ref{fig:compa} and \ref{fig:compb} as evidenced by the flatness of the curves for AEM, $T_2 = \infty$. However, AEM only partially protects the rate parameter against phase-flip error. This should not be surprising as phase error would persist even after the projection due to measurement at the end of a call to Grov. That is, phase error tends to accumulate in the superposition of the group register and is not corrected by the AEM strategy. Still looking at the SEM results, we see that the algorithm is altogether less susceptible to phase-flip error.   

The protection of the rate parameter by AEM is important as it means our choice of the oracle budget parameter is less dependent on knowing the rate of error. However, the rate parameter is no-longer directly proportional to the run-time of the algorithm as errored calls to Grov are not counted against the oracle budget. Thus we have to evaluate whether the total run-time is better or worse under AEM, which not only includes the errored calls, but also includes the time to perform the measurements. Fig.~\ref{fig:compc} and \ref{fig:compd} plots the average run-time as a function of either $T_1$  or $T_2$ for a fixed, large value of the other parameter. By average run-time, we mean the average over all trials of the total run-time (to find the correct answer) of the quantum computation cycles of the algorithm, including all measurements and gates, in units of the SQGT. This does not include time to perform the classical computation cycles of the algorithm\footnote{The variability in the average run-time is maximal as the time to reach the minimum can be zero if $v$ happens to be the minimum. For this reason, we give no error bars on the average run-time.}. From this figure, we see that AEM does not bloat the run-time for bit-flip error and as desired, significantly decreases the run-time for small $T_1$ times. It also only adds a modest, roughly constant increase for phase-flip error. Note that for higher $T_1$ and $T_2$ there is a cross-over where SEM has a smaller average run-time. This is due to the additional time needed to perform the measurements, which is only a constant time increase for each call to Grov.

From this analysis, we see that AEM is always preferred over SEM as it both protects the rate parameter and decreases the run-time except for when coherence times are sufficiently high, in which case its cost is only a constant for each call to Grov. 

\begin{figure*}[t]
\centering
\begin{tabular}{cc}
\subfloat[Rate Parameter \label{fig:scalea}]{\includegraphics[scale=.55]{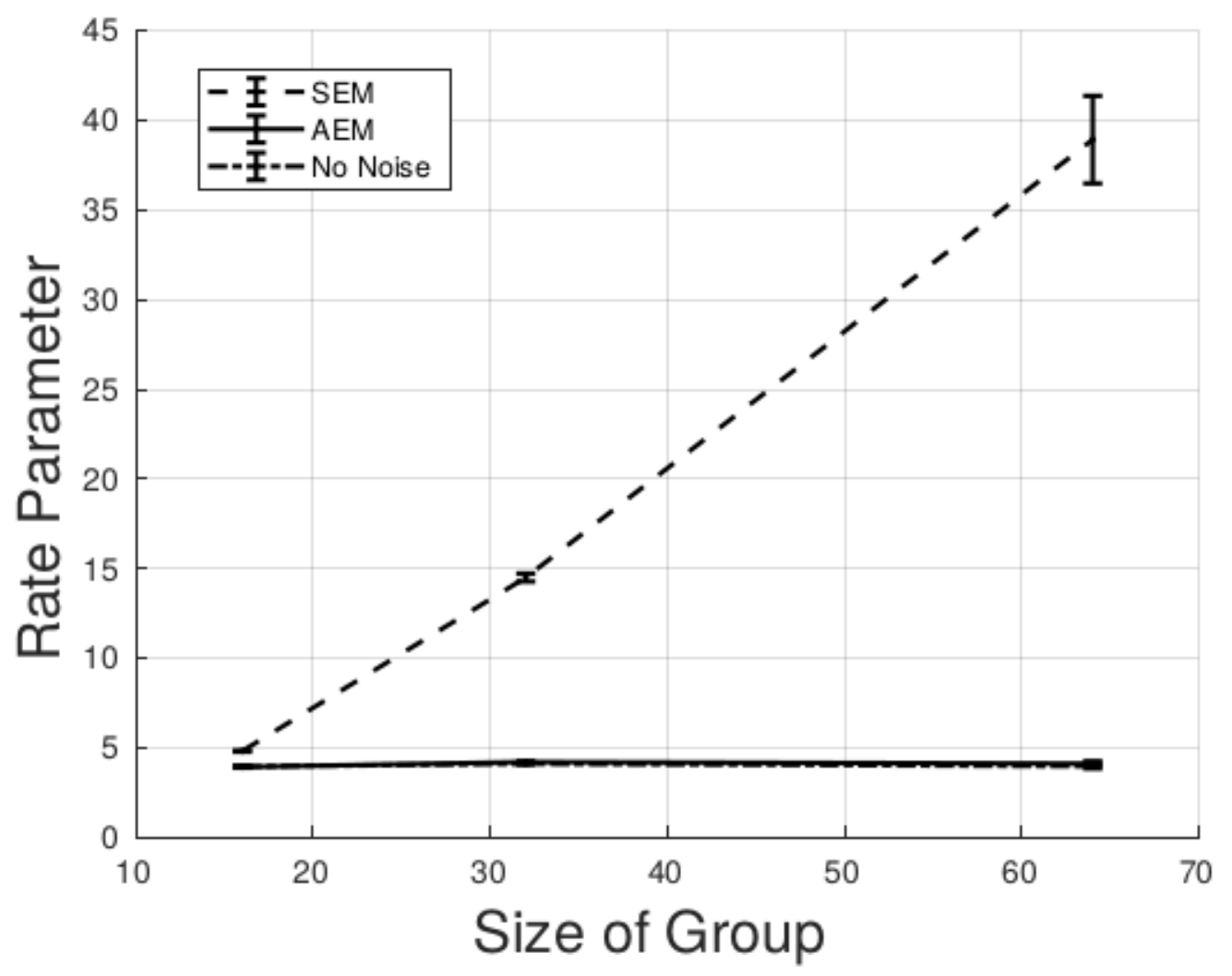}}&
\subfloat[Average Run-time\label{fig:scaleb}]{\includegraphics[scale=.55]{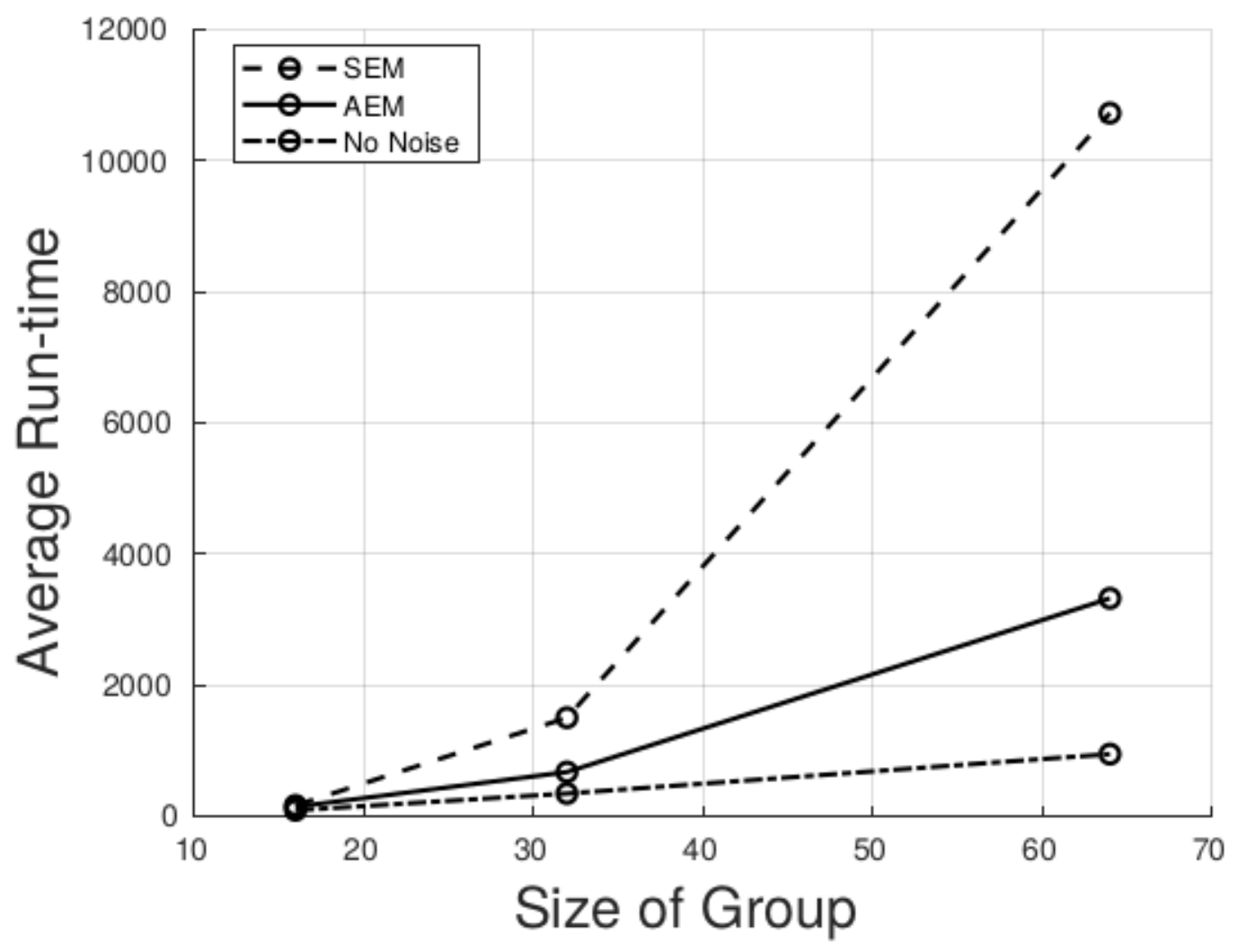}}
\end{tabular}
\caption{Simulation plots for rate parameter and average run-time using the realistic parameters $T_1 =T_2 = 700$ SQGTs. The rate parameter for AEM and no-noise are almost indistinguishable on this scale. Average run-time is also measured in units of the SQGT. The number of trials used is $4000$ for group sizes 16 and 32, and $500$ for group size 64. }\label{fig:scale}
\end{figure*}    

\subsection{Reducing Phase-flip Error}

AEM is effective against bit-flip error, but less so for phase-flip error. Even though the algorithm is less susceptible to this kind of error, it is worth considering a method for reducing phase-flip error. This can be achieved using simple fault-tolerant methods. As we are only looking to correct one channel of error, we can use simple, essentially classical fault-tolerant error-correcting codes such as a repetition code \cite{Terhal2015}. It should be sufficient to use an error-correcting code on the group register only to reduce the qubit overhead. With enough physical qubits to form robust logical qubits, we could achieve an effective $T_2 \sim \infty$ in which case AEM should fully protect the rate parameter.

\subsection{Simulation for Realistic Hardware}\label{sec:realsim}

AEM Gmin requires interaction between quantum and classical instructions, but unlike similar hybrid computations such as decoding an error-correcting code or variational eigensolver (VQE), the classical computation cycles are simple and should not take a significant amount of time between coherent quantum steps. Thus AEM Gmin could stand as a good test of real-time hybrid quantum-classical computation. For this reason, we simulate AEM Gmin with realistic $T_1, T_2$ times using the addition group of sizes $n = \log |G|= 4, 5$ and $6$.  To increase the chances of a successful run, we use the maximum number of ancilla qubits to reduce the depth of the circuit. So the total qubits used is $3n+(n-2)= 4n-2$, or $14, 18$ and $22$, respectively, for our cases. Methods for using the ancilla to reduce the depth are give in Appendix~\ref{redux}. We used $T_1 =T_2 =700$ SQGTs which are extracted from Ref.~\cite{O'Brien2017} for superconducting qubits. 

Fig.~\ref{fig:scale} plots the rate parameter and average run-time for AEM Gmin as well as SEM Gmin and no noise Gmin which are included for comparison. For these realistic hardware parameters, we see that the rate parameter is well-protected by AEM, and the increase in run-time over no-noise conditions is still within reason, whereas the time for SEM is beyond a reasonable run-time. When observing the simulation in real-time, we recognize for $n =6$ the probability of failure for a single oracle call is high, implying that a test of any larger groups would require an increase in the $T_1$ and $T_2$ times as argued in Section~\ref{sec:scale}. 

\section{Conclusions}

In this work, we have identified a new application for the Grover minimization algorithm, and provided a full quantum solution for the problem. Since Grover's search often comes with the caveat of not having an efficiently implementable oracle, our work is notable for finding a practical use for Grover's algorithm as the oracle is expected to scale poly-logarithmically with the size of the group. We have discussed both the structure of the algorithm and refinements to the original version, as well as a full gate decomposition for the simplest group given by modular addition. We discussed how we can leverage the intermediate measurement steps to mitigate the effects of error, increasing the likelihood of the algorithm being useful in the NISQ era.

In addition to being a sub-routine in classical exact diagonalization, our algorithm could also be called by a larger quantum algorithm which is performing a simulation of a many-body quantum system using symmetry-adapted basis states.

The algorithm discussed is far more general than what has been presented here. We achieve a reasonably sized oracle by leveraging the structure of the group, whereas the unstructured nature of the search is encapsulated in the arbitrary labeling of positions/basis states. Similarly, we can envisage using Gmin to find/prepare the ground state of some Hamiltonian. In such a case, one leverages the structure of Hamiltonian dynamics by replacing the group action operator with phase estimation. We hope to explore this more in future work.

The error mitigation scheme we have designed is also likely to be generally applicable to oracles using ancilla qubits, and thus could be used in a much wider context to improve the accuracy of quantum oracles.

\section{Acknowledgements}
The authors would like to thank Jim Held, Justin Hogaboam, Anne Matsuura,  and Xiang Zou for useful discussion. ATS would also like to thank Rahul M. Nandkishore. %\noteAS{How should I acknowledge Intel for funding?} \noteSJ{I added Intel affiliation for you since that's where you were when most of the research was done.}

\appendix

%\section{Explanation of the Symmetry-adapted Basis States}\label{ap:sym}

\section{Deriving a Tighter Lower Bound for the Oracle Budget}\label{derive}

In this appendix, we derive a tighter lower bound on the oracle budget. We follow the exact method used in Ref. \cite{Durr1996} but use a tighter bound from Ref. \cite{Boyer1998} for the average number of oracle calls for Gsun to find the solution to a search among $k$ marked elements. In particular, we use the exact expression for number of calls to reach the critical stage of the algorithm, with which we achieve a bound for Gsun of $\frac{9}{4}\frac{N}{\sqrt{k(N-k)}}$ ( whereas Ref. \cite{Durr1996} used $\frac{9}{2}\sqrt{\frac{N}{k}}$). Taking Lemma 1 from Ref. \cite{Durr1996} for granted, we follow the procedure for Lemma 2 using this tighter bound to find that the average number of oracle calls to reach the minimum is bound above by
\begin{align}
&\sum_{k=1}^N \frac{1}{k+1}\left(\frac{9}{4}\frac{N}{\sqrt{k(N-k)}}\right) \nonumber \\
 &= \frac{9 N}{4} \left( \frac{1}{2\sqrt{N-1}} + \sum_{k=2}^{N-1} \frac{1}{k+1} \frac{1}{\sqrt{(N-k) k}} \right).
\end{align}
We can approximate the sum using an integral as an upper bound,
\begin{align}
&\sum_{k=2}^{N-1} \frac{1}{k+1} \frac{1}{\sqrt{(N-k) k}}< \sum_{k=2}^{N-1} \frac{1}{k} \frac{1}{\sqrt{(N-k) k}} \nonumber \\
&< \int_1^{N-1} \frac{dk}{k} \frac{1}{\sqrt{(N-k) k}} \nonumber \\
 &= \left[-\frac{2}{N} \sqrt{\frac{(N-k)}{k}} \right]_{1}^{N-1} \nonumber \\
&=\frac{2}{N}\sqrt{N-1}\left(1  - \frac{1}{N-1}\right)
\end{align}
Ignoring the $\mc O\left(N^{-\frac{3}{2}}\right)$ term we find our upper bound is 
\begin{align}
\frac{9N}{8 \sqrt{N-1}} + \frac{9}{2} \sqrt{N-1} \sim \frac{45}{8} \sqrt{N}.
\end{align}

\section{Reducing the Cost of PhComp and $\hat G_{\text{add}}^N$}\label{redux}

\begin{figure*}
\centering
\includegraphics[scale=.7]{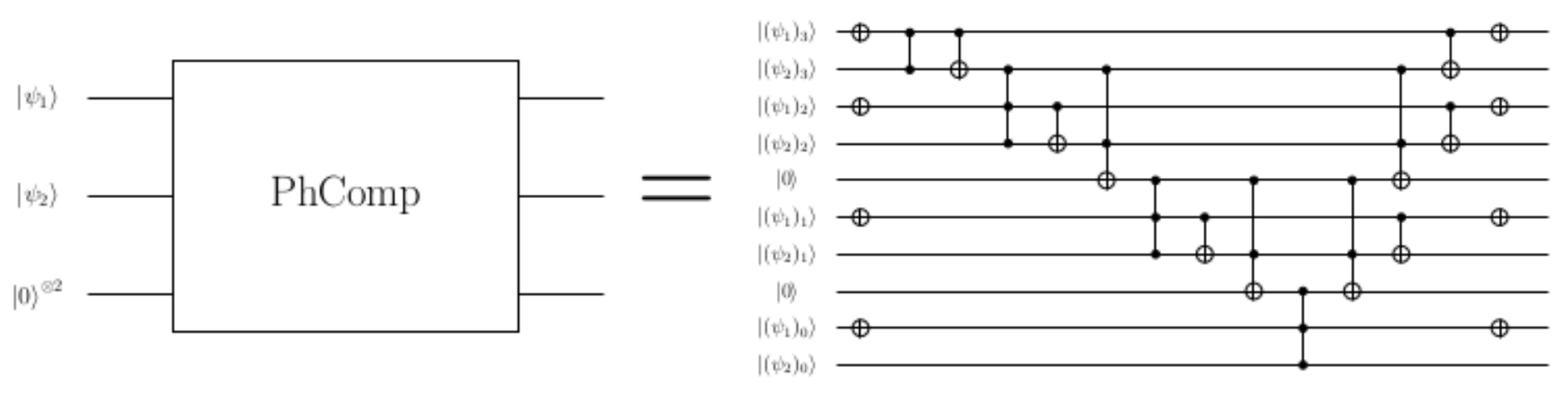}
\caption{Example of an $n=4$ PhComp circuit which uses two additional ancilla to reduce the number gates needed for implementation.}\label{phcompredux}
\end{figure*}

\begin{figure*}
\centering
\includegraphics[scale=.8]{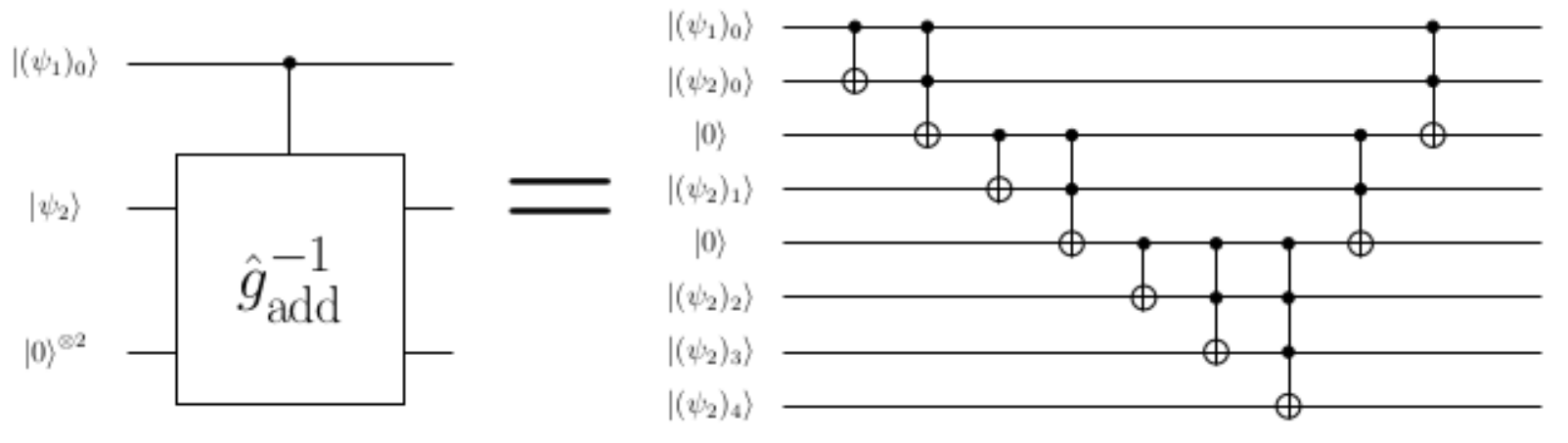}
\caption{Example of the zeroth bit part of $n=5$, $\left(\hat G_{\text{add}}^{2^n}\right)^{-1}$ circuit which uses two additional ancilla to reduce the number gates needed for implementation. We show the inverse as the method for using the ancilla is more clear. $G_{\text{add}}^{2^n}$ is then given by reversing the order of these gates. }\label{addredux}
\end{figure*}

To reduce the cost of PhComp, we avoid re-calculating the AND of (continue) bits i.e. remove the multi-control Z gates. This is done by storing the AND between two (continue) bits in an ancilla initialized in the zero computational state. We then pass this down the circuit as shown in Fig.~\ref{phcompredux}. The most significant and least significant bits do not benefit from having an ancilla, so we can use any number of ancilla up to $\log|V|-2$. For the maximum number, the cost of PhComp goes as $\mc C(\text{PhComp})\sim \mc O(\log|V|)$. A similar method can be used to reduce the cost of $\hat G_{\text{add}}^{N}$ as shown in Fig.~\ref{addredux}. With the maximum number of ancilla, which is again $\log|V|-2$, this reduces the cost of $\hat G_{\text{add}}^{N}$ to $\mc C(\hat G_{\text{add}}^{N})\sim \mc O(\log|G|)$. The resulting adder is on par with the ripple-carry adder from Ref.~\cite{Cuccaro2004}, but uses far more qubits. We include our version here to demonstrate that ancilla can be useful for reducing the group action operator. Furthermore, the ancilla can be shared between PhComp and the group action operator and measured along with the position registers in the AEM scheme. This was done for the data in Fig.~\ref{fig:scale}.

\section{Derivation of AEM Performance} \label{ap:AEMper}

In this appendix, we derive an estimate of the performance for AEM Gmin using a simple error model. Importantly, the analysis includes the measure-and-check strategy.

Let $\tilde{\mc G}$ be the noisy Grover call quantum channel. We make the assumption that we can decompose $\tilde{\mc G}$  as 
\begin{align}
\tilde{\mc G}(\rho) = \sigma \mc G(\rho) + (1-\sigma) \mc E(\rho),
\end{align}
for some $\sigma \in [0,1]$, where $\mc G$ is the noise-less Grover call quantum channel and $\mc E$ is some error channel. In  this version of AEM, we conditionally call $\tilde{\mc G}$ based upon the outcome of measuring the correct values in the position registers after the previous call to $\tilde{\mc G}$. Let $\mc P_C(\rho)= P_{v,v_{\text{best}}} \rho P_{v,v_{\text{best}}}$ be the channel which projects onto the correct computational basis states in the position registers and $\mc P_E(\rho)=\sum_{(u_1,u_2) \neq (v, v_{\text{best}})} P_{u_1, u_2} \rho P_{u_1, u_2}$ be the projection channel onto all incorrect basis states. We then model an AEM call to noisy Grover as
\begin{align}
\tilde{\mc G}_{\text{AEM}}(\rho)= \tilde{\mc G} \mc P_C(\rho) + \mc P_E(\rho).
\end{align}
We simplify the error channel by considering a model such that
\begin{subequations}
\begin{align}
\mc P_C \mc E(\rho)=& \mc P_C (\rho_{\text{mix}}), \\
\mc P_E \mc E(\rho)=& e  \mc P_E \mc F(\rho) + (1-e) \mc P_E (\rho_{\text{mix}}),  
\end{align}
\end{subequations}
where $e \in [0,1]$  and $\mc F$ is some quantum channel which only acts non-trivially on the position registers. For concreteness, we can think of $\mc F$ as some channel that applies an arbitrary string of Pauli $X$ operators with some probability, but the exact form does not matter for our purposes. $\rho_{\text{mix}}$ is the mixed state for the entire system. We interpret this error model as saying an error with the correct values in the position registers is effectively a maximally mixed state, and an error with the incorrect values in the position registers is such that it only affects those registers with some probability $e$ and is otherwise maximally mixed. That we take $e$ to be some value other than $0$ is informed by the fact that measure-and-check is numerically shown to significantly increase the probability of success.
   
Now suppose we apply $p$ AEM noisy Grover calls to the initial state,
\begin{align}
\rho_{\text{init}}= \outerp{s}{s} \otimes \outerp{v}{v} \otimes \outerp{v_{\text{best}}}{v_{\text{best}}},
\end{align}
followed by a final measurement of the position registers so that our final state is
\begin{widetext}
\begin{align}\label{eq:final1}
\rho_{\text{final}}= (\mc P_C + \mc P_E) \left(\tilde{\mc G}_{\text{AEM}}\right)^p (\rho_{\text{init}}) = \left(\mc P_C \tilde{\mc G}\right)^p  (\rho_{\text{init}}) + \mc P_E \sum_{n=1}^{p}\left( \tilde{\mc G} \mc P_C\right)^n  (\rho_{\text{init}}),
\end{align}
\end{widetext}
where we use the fact that $\mc P_C (\rho_{\text{init}})=  (\rho_{\text{init}})$, $\mc P_E  (\rho_{\text{init}}) = 0$ and $\mc P_E \mc P_C =0$. Once we substitute our error model into the above expression, we have several terms which are proportional to $\rho_{\text{mix}}$, noting that $\mc G(\rho_{\text{mix}}) = \mc F(\rho_{\text{mix}}) = \rho_{\text{mix}}$. These terms are sub-leading as their contribution to the final probability of success goes as $\frac{1}{N}$, so we collect all such terms in the set $\mc O(\rho_{\text{mix}})$. We then expand the errored terms in Eq. \eqref{eq:final1},
\begin{widetext}
\begin{align}
 \mc P_E \sum_{n=1}^{p}\left( \tilde{\mc G} \mc P_C\right)^n  (\rho_{\text{init}})=&  \mc P_E\tilde{\mc G} \sum_{n=0}^{p-1}\left(  \mc P_C \tilde{\mc G}\right)^n  (\rho_{\text{init}})
 = (1-\sigma) e \mc P_E \mc F \sum_{n=0}^{p-1} \sigma^n\left(  \mc P_C \mc G \right)^n  (\rho_{\text{init}}) + \mc O(\rho_{\text{mix}}),
\end{align}
\end{widetext}
where we are using $\mc P_E \mc G (\mc P_C \mc G)^n (\rho_{\text{init}}) =0$ as $\mc G$ acts as the identity on the position registers. Now suppose $\mc P_{\text{sol}}(\rho) = P_{\text{sol}}\rho P_{\text{sol}}$ is the projection channel for the solution space of the search. We then use the known exact solution for noise-less Grover search,
\begin{align}
\tr&\left( \mc P_{\text{sol}} \left(\mc P_C \mc G\right)^n (\rho_{\text{init}})\right)\nonumber \\
&=\tr\left( \mc P_{\text{sol}} \mc F \left(\mc P_C \mc G\right)^n(\rho_{\text{init}})\right)\nonumber\\
&= \sin^2((2n+1)\theta),
\end{align}
where $\theta$ is defined by $\sin^2 \theta= \frac{m}{N}$ for $m$ marked elements. Note we can apply $\mc F$ in the second equality as it only acts on the position registers and not the group register, i.e. the search space. Ignoring the $\mc O(\rho_{\text{mix}})$ terms, we can bound our success probability as
\begin{widetext}
\begin{align}
P_\text{success}^{(p)} = \tr \left(\mc P_{\text{sol}}( \rho_{\text{final}})\right) \geq \sigma^p \sin^2((2p+1)\theta) + (1-\sigma) e \sum_{n=0}^{p-1} \sigma^n\sin^2((2n+1)\theta).
\end{align}
\end{widetext}
The first term represents the probability of success when no error in the AEM scheme is detected and the other terms represent the probability of success when we measure the group register after an error is found at the $n^{th}$ Grover step. Using geometric series identities, we can perform the sum to find that
\begin{widetext}
\begin{align}\label{eq:sum}
 (1-\sigma)\sum_{n=0}^{p-1} \sigma^n&\sin^2((2n+1)\theta)\nonumber \\
&= \frac{(1-\sigma)}{2}\left(\frac{1-\sigma^p}{1-\sigma} -\frac{(1-\sigma)\cos(2\theta) - \sigma^p\cos(2(2p+1)\theta) + \sigma^{p+1} \cos(2(2p-1)\theta)}{1+\sigma^2-2 \sigma \cos(4 \theta)}\right) \nonumber \\
=& \frac{1-\sigma^p}{2}\left(1 - \frac{1}{1+\frac{4\sigma \sin^2\left(2\theta\right)}{(1-\sigma)^2}}\right)\nonumber \\
 &+ \frac{1}{1+\frac{4\sigma \sin^2\left(2\theta\right)}{(1-\sigma)^2}}\left(\sin^2(\theta) -\frac{ \sigma^p\left(\sin^2((2p+1)\theta) - \sigma\sin^2((2p-1)\theta)\right)}{1-\sigma}\right)
\end{align}
\end{widetext}

To simplify this expression, consider the case when $m =1$ and $N\gg 1$. In the denominator for both terms, we have the expression $\frac{4\sigma \sin^2\left(2\theta\right)}{(1-\sigma)^2}$, where care has to be taken as we have competing limits as $N\to \infty$, when assuming
\begin{align}
\sigma\sim \exp\left(-\frac{\mc C(\text{Grov})}{\braket{t}}\right),
\end{align}
 for the coherence time $\braket{t}$, which we also assume is a monotonically increasing function of $N$. %\red{Why do we specifically need this assumption on $\braket{t}$ here?}.
 Thus to lowest order in $\frac{1}{N}$ and using $\sin^2(2\theta) = 4\sin^2(\theta)\cos^2(\theta) = \frac{4}{N} +\mc O(\frac{1}{N^2})$, we find that
\begin{align}
\frac{4\sigma \sin^2\left(2\theta\right)}{(1-\sigma)^2} \sim 4 \left(\frac{4}{N}\right)\left(\frac{\braket{t}}{\mc C(\text{Grov})}\right)^2 \equiv \delta^2.
\end{align}
Looking back at Eq.~\eqref{eq:sum}, $\delta = \mc O(1)$ for AEM with measure-and-check to significantly increase the probability of success. %\red{Do we really need this assumption on $\delta$ here? },
So in terms of $\delta$, the coherence time goes as
\begin{align}
\braket{t} \sim\frac{\delta}{4}\mc C(\text{Grov})\sqrt{N}.
\end{align}
To give a final expression for the probability of success, we make a few approximations. First we use $\sin^2\left( (2p+1)\theta \right) = \sin^2\left( (2p-1)\theta \right) + \mc O\left(\frac{1}{N}\right)$ and likewise, we ignore the term in Eq. \eqref{eq:sum} which goes as $\sim \sin^2\theta= \frac{1}{N}$. %Finally we note that $e$ should have a similar exponential form as $s$ but as $\braket{t} \sim \sqrt{N}$, we can reasonable take $e \to 1$ as $N\to \infty$ %\red{Maybe you should call $e$ something else since in the text we use $e^{-1}$ etc. to be the exponential.}.
Our probability of success is then asymptotically 
%
%\begin{widetext}
\begin{align}
P_{\text{success}}^{(p)}\sim&
\left(1-\frac{e}{1+ \delta^2}\right)\sigma^p \sin^2\left((2p+1)\theta\right) \nonumber \\
&+\frac{e\delta^2}{1+ \delta^2} \frac{(1-\sigma^p)}{2} ,
\end{align}
%\end{widetext}
%
where $\sigma \sim \exp\left(-\frac{4}{\delta \sqrt{N}}\right)$. When $e=1$, i.e. the most optimistic case, this reduces to Eq. \eqref{eq:aemprob}.

\section{Details of the Noisy Simulation}\label{detail}

In this appendix, we discuss some of the details of the noisy simulation. 
Relative gate times are extracted from Ref.~\cite{O'Brien2017} which uses data for superconducting qubits. All single qubit gate times (SQGT) are assumed to be equal and all other simulation times are measured in units of this time. All two-qubit gates are assumed to be twice the SQGT and all gates are decomposed into one- and two-qubit gates. The Intel-QS does not have a feature to simulate measurements, so our source code has been altered to include measurement simulation capabilities. A Mersenne twist random number generator is added specifically to simulate the probabilistic nature of quantum measurement. Furthermore, a measurement time of 10 SQGTs is added to simulate the accumulation of error that would occur in a real system while a measurement is being performed. We do not consider the possibility of error in the measured value as compared to the resulting quantum state though this is an important source of error to consider in a real system. Finally, the method by which the Intel-QS accounts for the time between gate action has been altered to include parallelization. A sequence of gates with disjoint support on the qubits is assumed to be applied in parallel in which case time is only incremented by the largest gate time in that sequence. No error is accumulated during classical computation cycles, though this is an important source of error to consider for real systems. All these considerations are used to calculate the total run-time for a single trial of Gmin.

We also note that currently we do not make use of a compiler to reduce the number of gates. Therefore the error rates for all simulations are higher than they would be if we used such an optimizing software.

\bibliographystyle{unsrt}
\bibliography{citations}

\end{document}